\documentclass[lettersize,journal]{IEEEtran}
\usepackage{amsmath, amssymb, amsthm, amsfonts}
\usepackage[hidelinks]{hyperref}
\usepackage{graphicx}
\usepackage{booktabs}
\usepackage{bm}
\usepackage{cite}
\usepackage{color}
\usepackage{mathdots}
\usepackage{optidef}
\usepackage{soul}
\usepackage[caption=false,font=normalsize,labelfont=sf,textfont=sf]{subfig}

\usepackage{orcidlink}

\usepackage{textcomp}

\def\BibTeX{{\rm B\kern-.05em{\sc i\kern-.025em b}\kern-.08em
		T\kern-.1667em\lower.7ex\hbox{E}\kern-.125emX}}
\AtBeginDocument{\definecolor{ojcolor}{cmyk}{0.93,0.59,0.15,0.02}}

\newcommand{\unitvec}[1]{\hat{#1}}
\newcommand{\imag}{j}

\newcommand{\conj}[1]{\overline{#1}}
\newcommand{\herm}{\mathsf{H}}

\newcommand{\complex}{\mathbb{C}}
\newcommand{\real}{\mathbb{R}}

\newcommand{\integer}{\mathbb{Z}}

\newcommand{\unitsphere}{\mathbb{S}^2}
\usepackage{tikz}
\usepackage{pgfplots}
\pgfplotsset{compat=newest}
\usetikzlibrary{calc, positioning, backgrounds, arrows, arrows.meta, bending, decorations.markings, decorations.text}
\usepgfplotslibrary{groupplots}

\pgfplotsset{
	layers/my layer set/.define layer set={
		bg,
		main,
		foreground
	}{
	},
	set layers=my layer set,
}

\colorlet{labelcolor}{white!15!black}
\colorlet{ticklabelcolor}{white!45!black}
\colorlet{gridcolor}{white!85!black}

\colorlet{legendbordercolor}{white!100!black}

\pgfmathsetmacro{\groupplotsep}{17}
\pgfmathsetmacro{\timeplotheight}{2.8}
\pgfmathsetmacro{\timeplotwidth}{7.8}
\pgfplotsset{time-plot/.style={
		scale only axis,
		legend cell align={left},
		legend style={fill opacity=0, draw opacity=0, text opacity=1, draw=legendbordercolor, font=\scriptsize},
		tick align=inside,
		tick pos=left,
		xmajorgrids,
		ymajorgrids,
		grid style={gridcolor, line width=0.2pt, opacity=0.4},
		ticklabel style = {font=\scriptsize, ticklabelcolor,},
		y label style = {font=\footnotesize, labelcolor, rotate=-90, at=(ticklabel cs:1.1), anchor=west, inner sep=0pt},
		x label style = {font=\footnotesize, labelcolor},
		ytick style={draw=none},
		xtick style={draw=none},
		axis line style={draw=none},
		ylabel shift = -3 pt,
		xlabel near ticks,
}}

\pgfplotsset{position-plot/.style={
		scale only axis,
		axis equal image,
		legend cell align={left},
		legend style={fill opacity=0, draw opacity=1, text opacity=1, at={(0.03,0.97)}, anchor=north west, draw=legendbordercolor, font=\scriptsize},
		tick align=inside,
		grid style={gridcolor, line width=0.2pt, opacity=0.4},
		xmajorgrids,
		ymajorgrids,
		tick style={draw=none},
		ticklabel style = {font=\scriptsize, ticklabelcolor},
		y label style = {font=\footnotesize, labelcolor},
		x label style = {font=\footnotesize, labelcolor},
		xlabel = {x (m)},
		ylabel = {y (m)},
		axis line style = { draw = none },
}}

\pgfplotsset{soundfield-plot/.style={
		axis on top=true,
		scale only axis,
		axis equal image,
		axis line style = { draw = none },
		tick align = inside,
		tick pos=left,
		xtick style={color=black},
		ytick style={color=black},
		ticklabel style = {font=\scriptsize, ticklabelcolor},
		y label style = {font=\footnotesize, labelcolor},
		x label style = {font=\footnotesize, labelcolor},
		xlabel = {x (m)},
		ylabel = {y (m)},
		ylabel shift = -10 pt,
}}

\pgfplotsset{matrix-plot/.style={
		axis on top=true,
		scale only axis,
		axis equal image = true, 
		axis line style = { draw = none },
		tick align = inside,
		tick pos=left,
		xtick style={color=black},
		ytick style={color=black},
		ticklabel style = {font=\scriptsize, ticklabelcolor},
		y label style = {font=\footnotesize, labelcolor},
		x label style = {font=\footnotesize, labelcolor},
}}

\pgfplotsset{spectogram-plot/.style={
		axis on top=true,
		scale only axis,
		axis line style = { draw = none },
		tick align = inside,
		tick pos=left,
		xtick style={color=black},
		ytick style={color=black},
		ticklabel style = {font=\scriptsize, ticklabelcolor},
		y label style = {font=\footnotesize, labelcolor},
		x label style = {font=\footnotesize, labelcolor},
		ylabel shift = -10 pt,
		colorbar style={
			axis equal image = false,
			width=0.2*\pgfkeysvalueof{/pgfplots/parent axis width},
			height=0.8*\pgfkeysvalueof{/pgfplots/parent axis height},
			ylabel shift = 0 pt,
		},
}}

\pgfplotsset{
	micmarker/.style={
		semithick, 
		mark=*, 
		mark size=1.8, 
		only marks,
		draw=black,
		line width=0.5,
		mark options=solid,
}}

\pgfplotsset{
	srcmarker/.style={
		semithick, 
		mark=square*,
		mark size=3,
		only marks,
		draw=black, 
		line width=0.7,
		mark options=solid,
}}

\pgfmathsetmacro{\standardlinewidth}{1.5}
\pgfmathsetmacro{\offamount}{\standardlinewidth*1.7}
\pgfmathsetmacro{\standardlineopacity}{0.9}
\newcommand{\linejoinchoice}{round}

\pgfplotsset{
	line0/.style={
		color0, 
		opacity=\standardlineopacity, 
		line cap=round, 
		line width = \standardlinewidth pt, 
		mark options={solid},
		line join=\linejoinchoice,
}}

\pgfplotsset{
	line1/.style={
		color1, 
		opacity=\standardlineopacity, 
		dash pattern=on 1pt off \offamount pt,
		line cap=round,
		line width = \standardlinewidth pt, 
		mark options={solid},
		line join=\linejoinchoice,
}}

\pgfplotsset{
	line2/.style={
		color2, 
		opacity=\standardlineopacity, 
		dash pattern=on 3.5pt off \offamount pt, 
		line cap=round,
		line width = \standardlinewidth pt, 
		mark options={solid},
		line join=\linejoinchoice,
}}
\pgfplotsset{
	line3/.style={
		color3, 
		opacity=\standardlineopacity, 
		dash pattern=on 6pt off \offamount pt,
		line cap=round,
		line width = \standardlinewidth pt, 
		mark options={solid},
		line join=\linejoinchoice,
}}
\pgfplotsset{
	line4/.style={
		color4, 
		opacity=\standardlineopacity, 
		dash pattern=on 8pt off \offamount pt,
		line cap=round,
		line width = \standardlinewidth pt, 
		mark options={solid},
		line join=\linejoinchoice,
}}
\pgfplotsset{
	line5/.style={
		color5, 
		opacity=\standardlineopacity, 
		dash pattern=on 10pt off \offamount pt,
		line cap=round,
		line width = \standardlinewidth pt, 
		mark options={solid},
		line join=\linejoinchoice,
}}
\pgfplotsset{
	line6/.style={
		color7, 
		opacity=\standardlineopacity, 
		dash pattern=on 12pt off \offamount pt,
		line cap=round,
		line width = \standardlinewidth pt, 
		mark options={solid},
		line join=\linejoinchoice,
}}

\pgfplotsset{
	linegrad0/.style={
		color0!20!white, 
		opacity=\standardlineopacity, 
		line cap=round, 
		line width = \standardlinewidth + 2.5 pt, 
		mark options={solid},
		line join=\linejoinchoice,
}}

\pgfplotsset{
	linegrad1/.style={
		color0!45!white, 
		opacity=\standardlineopacity, 
		line cap=round, 
		line width = \standardlinewidth pt, 
		mark options={solid},
		line join=\linejoinchoice,
}}

\pgfplotsset{
	linegrad2/.style={
		color0!80!white, 
		opacity=\standardlineopacity, 
		line cap=round, 
		line width = \standardlinewidth pt, 
		mark options={solid},
		line join=\linejoinchoice,
}}

\pgfplotsset{
	linegrad3/.style={
		color0!85!black, 
		opacity=\standardlineopacity, 
		line cap=round, 
		line width = \standardlinewidth pt, 
		mark options={solid},
		line join=\linejoinchoice,
}}

\pgfplotsset{
	linegrad4/.style={
		color0!55!black, 
		opacity=\standardlineopacity, 
		line cap=round, 
		line width = \standardlinewidth pt, 
		mark options={solid},
		line join=\linejoinchoice,
}}

\pgfplotsset{
	linegrad5/.style={
		color0!10!black, 
		opacity=\standardlineopacity, 
		line cap=round, 
		line width = \standardlinewidth pt, 
		mark options={solid},
		line join=\linejoinchoice,
}}

\definecolor{color0}{rgb}{0.00392156862745098, 0.45098039215686275, 0.6980392156862745}
\definecolor{color1}{rgb}{0.8705882352941177, 0.5607843137254902, 0.0196078431372549}
\definecolor{color2}{rgb}{0.00784313725490196, 0.6196078431372549, 0.45098039215686275}
\definecolor{color3}{rgb}{0.8352941176470589, 0.3686274509803922, 0.0}
\definecolor{color4}{rgb}{0.8, 0.47058823529411764, 0.7372549019607844}
\definecolor{color5}{rgb}{0.792156862745098, 0.5686274509803921, 0.3803921568627451}
\definecolor{color6}{rgb}{0.984313725490196, 0.6862745098039216, 0.8941176470588236}
\definecolor{color7}{rgb}{0.5803921568627451, 0.5803921568627451, 0.5803921568627451}
\definecolor{color8}{rgb}{0.9254901960784314, 0.8823529411764706, 0.2}
\definecolor{color9}{rgb}{0.33725490196078434, 0.7058823529411765, 0.9137254901960784}

\colorlet{color0grad0}{color0!20!white}
\colorlet{color0grad1}{color0!45!white}
\colorlet{color0grad2}{color0!80!white}
\colorlet{color0grad3}{color0!85!black}
\colorlet{color0grad4}{color0!55!black}
\colorlet{color0grad5}{color0!10!black}

\colorlet{color3grad0}{color3!20!white}
\colorlet{color3grad1}{color3!45!white}
\colorlet{color3grad2}{color3!80!white}
\colorlet{color3grad3}{color3!85!black}
\colorlet{color3grad4}{color3!55!black}
\colorlet{color3grad5}{color3!10!black}

\usepackage{etoolbox}
\usepackage{siunitx}
\DeclareSIUnit\octave{oct}
\robustify\bfseries

\usepackage[acronym]{glossaries}
\newacronym{rosance}{ROSANCE}{RObust Signal And Noise Covariance Estimator}
\newacronym{ces}{CES}{complex elliptically symmetric}
\newacronym{scm}{SCM}{sample covariance matrix}
\newacronym{dof}{DoF}{degrees of freedom}

\newacronym{dft}{DFT}{discrete Fourier transform}
\newacronym{fft}{FFT}{fast Fourier transform}

\newacronym{rir}{RIR}{room impulse response}
\newacronym{iir}{IIR}{infinite impulse response}
\newacronym{fir}{FIR}{finite impulse response}

\newacronym{szc}{SZC}{sound zone control}
\newacronym{vast}{VAST}{variable span trade-off filter}
\newacronym{acc}{ACC}{acoustic contrast control}
\newacronym{pm}{PM}{pressure matching}

\newacronym{qos}{QoS}{quality of service}
\newacronym{fpi}{FPI}{fixed point iteration}

\newacronym{mse}{MSE}{mean square error}
\newacronym{nmse}{NMSE}{normalized mean square error}
\newacronym{ac}{AC}{acoustic contrast}
\newacronym{sd}{SD}{signal distortion}
\newacronym{re}{RE}{residual energy}
\newacronym{sinr}{SINR}{signal-to-interference-plus-noise ratio}
\newacronym{snr}{SNR}{signal-to-noise ratio}

\newacronym{evd}{EVD}{eigenvalue decomposition}
\newacronym{gevd}{GEVD}{generalized eigenvalue decomposition}
\newacronym{pevd}{PEVD}{polynomial eigenvalue decomposition}
\newacronym{pgevd}{PGEVD}{polynomial generalized eigenvalue decomposition}
\newacronym{svd}{SVD}{singular value decomposition}
\newacronym{mwf}{MWF}{multichannel Wiener filter}

\newacronym{ola}{OLA}{overlap-add}
\newacronym{ols}{OLS}{overlap-save}
\newacronym{wola}{WOLA}{weighted overlap-add}

\newacronym{lms}{LMS}{least mean squares}
\newacronym{nlms}{NLMS}{normalized least mean squares}
\newacronym{fxlms}{FxLMS}{filtered-x least mean squares}
\newacronym{rls}{RLS}{recursive least squares}

\newacronym{pseq}{PSEQ}{perfect sequence}
\newacronym{ops}{OPS}{orthogonal periodic sequence}
\newacronym{pps}{PPS}{perfect periodic sequence}

\newacronym{mnf}{MNF}{minimum noise fraction}
\newacronym{rkhs}{RKHS}{reproducing kernel Hilbert space}
\newacronym{krr}{KRR}{kernel ridge regression}
\newacronym{rff}{RFF}{random Fourier features}

\newacronym{psd}{PSD}{positive semi-definite}
\newacronym{sdr}{SDR}{semi-definite relaxation}
\newacronym{sdp}{SDP}{semi-definite program}
\newacronym{qcqp}{QCQP}{quadratically constrained quadratic program}

\makeatletter
\newcommand*{\centerfloat}{%
	\parindent \z@
	\leftskip \z@ \@plus 1fil \@minus \textwidth
	\rightskip\leftskip
	\parfillskip \z@skip}
\makeatother

\usepackage{soul}

\DeclareMathOperator{\linreg}{\mathcal{R}}
\newcommand{\rkhs}{\mathcal{H}}
\newcommand{\rkhstime}{\tilde{\mathcal{H}}}

\newcommand{\td}{\mathfrak{t}}
\newcommand{\fd}{\mathfrak{f}}

\newcommand{\fdmatrix}{\mathcal{B}(\fd)}
\newcommand{\tdmatrix}{\mathcal{B}(\td)}

\newcommand{\pwcoeff}[1]{\mathring{#1}}

\begin{document}

	\title{Sound field estimation with moving microphones using kernel ridge regression}
	
	\author{Jesper Brunnström \orcidlink{0000-0003-2946-1268}, Martin Bo Møller \orcidlink{0000-0002-7323-7266}, Jan Østergaard \orcidlink{0000-0002-3724-6114}, Shoichi Koyama \orcidlink{0000-0003-2283-0884}, Toon van Waterschoot \orcidlink{0000-0002-6323-7350}, and Marc Moonen \orcidlink{0000-0003-4461-0073}
	\thanks{This research work was carried out at the ESAT Laboratory of KU Leuven, in the frame of Research Council KU Leuven C14-21-0075 "A holistic approach to the design of integrated and distributed digital signal processing algorithms for audio and speech communication devices", and has received funding from the European Union's Horizon 2020 research and innovation programme under the Marie Skłodowska-Curie grant agreement No. 956369: 'Service-Oriented Ubiquitous Network-Driven Sound — SOUNDS'. The scientific responsibility is assumed by the authors.}
	\thanks{Jesper Brunnström, Toon van Waterschoot, and Marc Moonen are with the Department of Electrical Engineering (ESAT), STADIUS, KU Leuven, Leuven, Belgium (e-mail: \url{jesper.brunnstroem@kuleuven.be})}
	\thanks{Martin Bo Møller is with Bang \& Olufsen, Acoustics R\&D, Struer, Denmark} 
	\thanks{Jan Østergaard is with the Department of Electronic Systems, Aalborg University, Aalborg, Denmark} 
	\thanks{Shoichi Koyama is with the National Institute of Informatics, Tokyo 101-8430, Japan}
	}

	\maketitle
	\begin{abstract}
		Sound field estimation with moving microphones can increase flexibility, decrease measurement time, and reduce equipment constraints compared to using stationary microphones. In this paper a sound field estimation method based on \acrfull{krr} is proposed for moving microphones. The proposed \acrshort{krr} method is constructed using a discrete time continuous space sound field model based on the discrete Fourier transform and the Herglotz wave function. The proposed method allows for the inclusion of prior knowledge as a regularization penalty, similar to kernel-based methods with stationary microphones, which is novel for moving microphones. Using a directional weighting for the proposed method, the sound field estimates are improved, which is demonstrated on both simulated and real data. Due to the high computational cost of sound field estimation with moving microphones, an approximate \acrshort{krr} method is proposed, using \acrfull{rff} to approximate the kernel. The \acrshort{rff} method is shown to decrease computational cost while obtaining less accurate estimates compared to \acrshort{krr}, providing a trade-off between cost and performance.

	\end{abstract}
	\begin{IEEEkeywords}
		Sound field estimation, moving microphone, kernel ridge regression, reproducing kernel Hilbert space, random Fourier features
	\end{IEEEkeywords}

	\section{Introduction}
	\IEEEPARstart{T}{he} development of effective sound field estimation methods is crucial for many sound field control applications \cite{betlehemPersonal2015, zhangSurround2017}. Sound field estimation is difficult due to the large number of spatial samples required to perfectly reconstruct the sound field \cite{ajdlerPlenacoustic2006}. The sound field is therefore in practice spatially undersampled, and sound field estimation is performed by solving an ill-posed inverse problem using some type of prior knowledge as regularization.
	
	One ubiquitous assumption for sound field estimation is that the microphones are stationary during the measurement process. This assumption simplifies the derivations of the methods, but then requires that microphones are kept still. Estimating sound fields over large regions require many spatial samples, which either takes a long time to collect or requires expensive large microphone arrays. Using moving microphones allows for more flexible measurements, which can shorten the measurement time or relax equipment requirements. In addition, sound field estimation could be applied in situations where the microphones cannot be kept stationary.
	
	Of particular interest for sound field reproduction applications is estimating the \gls{rir} function associated with a specific source, which is therefore the focus of this paper. The \gls{rir} function is the sound field generated when the source signal is an impulse. It is impractical to use an impulse as a source signal in practice, so instead sine-sweep signals are commonly used \cite{farinaSimultaneous2000, mullerTransferfunction2001, antweilerNLMStype2008}, after which the received signal must be deconvolved to obtain a \gls{rir}. Using stationary microphones, the recorded signal for each microphone can first be deconvolved such that \glspl{rir} are obtained, followed by estimating \glspl{rir} at arbitrary positions through sound field interpolation. Using moving microphones, the deconvolution and sound field interpolation must be jointly solved. This both complicates the derivation of the estimation methods, and considerably increases their computational cost.  
	

	While no specific constraints on the microphone trajectory is imposed in this paper, the trajectory needs to be known. This corresponds to the requirement that the microphone positions must be known in sound field estimation with stationary microphones. In a similar way to how stationary microphone positions affects the sound field estimation performance \cite{ajdlerPlenacoustic2006, nishidaRegionrestricted2022, verburgOptimal2024}, the choice of trajectory also affects the estimation performance \cite{unnikrishnanSampling2013a, grochenigMinimal2015, katzbergCoherence2021}. Under some constraints on the source signal, the trajectory can be obtained from the recorded audio signal itself, which means that a separate tracking system is not necessarily required \cite{katzbergPositional2022}. If the obtained trajectory is not accurate, sound field estimation accuracy is degraded due to the Doppler shift \cite{katzbergDoppler2024}.
	

	There exists some methods for sound field estimation with moving microphones. The methods in \cite{ajdlerDynamic2007, hahnComparison2016, hahnContinuous2017, hahnSimultaneous2018} are suited to estimation on the one-dimensional trajectory along which the microphone moves, some of which are relying on the specific properties of a perfect periodic sequence \cite{antweilerPerfect1994, antweilerNLMStype2008} as source signal. The problem has also be approached as a sampling problem \cite{katzbergSoundfield2017}, as well as using compressive sensing to reduce measurement time \cite{katzbergCompressed2018}.



	It can be beneficial to include a sound field model incorporating known properties of sound, as is often done with sound field estimation with stationary microphones. This can be done through basis expansion, where the sound field is expressed as a superposition of position-dependent basis functions weighted by position-independent coefficients. Two of the most common basis functions are plane waves and spherical wave functions, both of which are analytical solutions to the homogenous Helmholtz equation \cite{williamsFourier1999}. Such basis expansions have been used with spherical wave functions in \cite{katzbergSpherical2021, brunnstromBayesian2025} and plane waves in \cite{verburgDynamic2025} for the moving microphone case. 
	
	A basis expansion is in general an approximate sound field model, because only a finite number of basis functions are used in order to maintain computational tractability. However, it is possible to use an infinite number of basis functions to represent the sound field, as was done in \cite{brunnstromBayesian2025} using a similar approach to what was used for stationary microphones in \cite{uenoSound2018, brunnstromBayesian2024}. It can be viewed as an implicit basis expansion, as the basis coefficients never has to be computed explicitly. Experiments have shown that such an exact representation is more robust and produces better estimates \cite{brunnstromBayesian2025, brunnstromExperimental2025} compared to a truncated basis expansion \cite{katzbergSpherical2021}. 
	
	The infinite-dimensional basis expansion method for stationary microphones in \cite{uenoSound2018} is equivalent to \gls{krr} \cite{uenoKernel2018}. However, the \gls{krr} method has been further developed to incorporate prior knowledge of the sound field \cite{uenoDirectionally2021, horiuchiKernel2021, ribeiroSound2024}, as well as taking other fundamental properties of the sound field into account \cite{ribeiroPhysicsconstrained2024, matsudaKernel2025}, significantly improving estimation performance. Therefore, it would be beneficial to develop \gls{krr} for the moving microphone case as well, as existing sound field estimation methods for moving microphones do not easily allow for the inclusion of prior knowledge of the sound field. However, the \gls{krr} methods for stationary microphones have been developed entirely in the frequency domain, making it non-trivial to adapt to the moving microphone case. 
	
	

	Recently, the \gls{krr} method for sound field estimation was extended to the discrete time-domain \cite{brunnstromTimedomainsubmitted}, which allows for the estimation of \glspl{rir}. This extension allows for time-domain characteristics of the sound field to be incorporated as prior knowledge, but also extends the class of problems that can be solved using the approach. One such problem is sound field estimation with moving microphones, which requires a time-domain model in order to perform the sound field interpolation and deconvolution jointly. 
	
	
	
	
	It is well known that the computational cost of \gls{krr} scales poorly with the number of data points, which for moving microphones is the number of time-domain samples measured. A method for reducing the computational cost of kernel methods is to use \gls{rff}, where a finite basis expansion is used to approximate the kernel estimate \cite{rahimiRandom2007, liuRandom2022}. From the proposed \gls{krr} method, in this paper \gls{rff} is developed for both stationary and moving microphones. \Gls{rff} in this case is a finite basis expansion onto a plane wave basis, as in \cite{verburgDynamic2025}, which demonstrates the relationship of finite plane wave basis expansion methods as an approximation of the proposed \gls{krr} method. This is similar to the relationship between the truncated spherical wave function expansion of \cite{katzbergSpherical2021} and the non-truncated expansion of \cite{brunnstromBayesian2025}. However, one benefit of \gls{rff} is that flexible regularization penalties can be derived in the \gls{rkhs} domain, which then translates to \gls{rff} as well.  
	
	
	

	To summarize, in this paper a kernel ridge regression approach to moving microphone sound field estimation is proposed, which in its basic form is equivalent to the Bayesian estimation method proposed in \cite{brunnstromBayesian2025}. The proposed \gls{krr} approach allows for the inclusion of prior knowledge through regularization schemes that have been developed in \cite{uenoDirectionally2021, horiuchiKernel2021, ribeiroSound2024, brunnstromSpatial2025, brunnstromTimedomainsubmitted}. The paper follows the development of a time-domain sound field estimation method based on kernel ridge regression developed in \cite{brunnstromTimedomainsubmitted}, and extends the approach to moving microphones. The proposed method is evaluated on both simulated data and real data from \cite{brunnstromExperimental2025}, demonstrating its effectiveness compared to existing methods.

	\subsection{Notation}
	The vector space of $N$-dimensional real vectors and of $N$-dimensional complex vectors are denoted by $\real^{N}$ and $\complex^{N}$ respectively. The unit sphere in three dimensions is denoted by $\unitsphere$, defined as $\{\bm{x} \;\vert \;\lVert \bm{x} \rVert = 1, \bm{x} \in \real^3\}$. The complex conjugate of $a$ is $\conj{a}$, the transpose of $\bm{A}$ is $\bm{A}^\top$, and the Hermitian transpose of $\bm{A}$ is $\bm{A}^\herm$. The imaginary unit is $\imag^2 = -1$. The same notation is used for linear operators as matrix multiplication, where the operator is applied to the variable to the right of it, e.g. in the expression $T a$, the variable $a$ is the argument of the linear operator $T$. The adjoint of a linear operator $T : \mathcal{H} \rightarrow \mathcal{H}'$ between Hilbert spaces $\mathcal{H}$ and $\mathcal{H}'$, is denoted by $T^{*} : \mathcal{H}' \rightarrow \mathcal{H}$. The selection of an element with index $l$ from $\bm{a}$ is denoted by $(\bm{a})_l$.
	
	
	
	
	\section{Problem statement}\label{sec:problem-statement}
	The task considered in this paper is to reconstruct \glspl{rir} at all positions within a region of interest, using measurements collected by a continuously moving microphone. The region of interest $\Omega \subset \real^3$ is a simply connected source-free region within which the microphone collects measurements. Outside the region $\Omega$ there is a sound source of interest which emits a known signal $\phi$, e.g. a loudspeaker using a predetermined training signal. 
	
	
	The acoustic response of the source and the room is assumed to be constant over the duration of the measurement process. The response measured by an ideal stationary omnidirectional microphone at the position $\bm{r} \in \Omega$ can be modelled as a constant $L$-length \gls{fir} filter, the impulse response of which is referred to as a \gls{rir}. The \glspl{rir} associated with all positions $\bm{r} \in \Omega$ can then be represented by a \gls{rir} function $\tilde{\bm{u}} : \Omega \rightarrow \td$, where $\td$ is the Hilbert space of $L$-length sequences with inner product 
	\begin{equation}
		\langle \tilde{\bm{u}}, \tilde{\bm{v}} \rangle_{\td} = \tilde{\bm{v}}^\top \tilde{\bm{u}}
		\label{eq:td-inner-product}
	\end{equation}
	which is equivalent to the standard Euclidean vector space $\real^L$. 
	
	
	Now consider an omnidirectional microphone moving continuously inside $\Omega$. The microphone samples the signal at a sampling rate of $f_s$, obtaining the time-domain signal $p(n) \in \real$, where $n \in \integer$ denotes the discrete time index. The position of the microphone is synchronously recorded, giving a sequence of positions $\bm{r}_0, \dots, \bm{r}_{N-1}$ representing the trajectory along which the microphone moves. The source signal is similarly sampled, giving the known signal $\phi(n) \in \real$.

	Given the presence of measurement noise $s(n) \in \real$, the obtained data can then be expressed as 
	\begin{equation}
		p(n) = \langle \tilde{\bm{u}}(\bm{r}_n), \bm{\phi}(n) \rangle_{\td} + s(n) 
		\label{eq:data-model}
	\end{equation}
	for time indices $n = 0, \dots, N-1$, and $\bm{\phi}(n) = (\phi(n), \dots, \phi(n-L+1)) \in \td$ containing the last $L$ samples of the source signal. The task is then to estimate the sound field function $\tilde{\bm{u}}$ from the available data. Note that if the source signals are chosen appropriately, the \gls{rir} functions associated with multiple sources can be simultaneously estimated \cite{hahnSimultaneous2018} \cite[Sec.IV.E]{brunnstromBayesian2025}.

	
	


	\section{Reproducing kernel Hilbert space}\label{sec:sound-field-model}
	The \gls{rir} function will be estimated using \gls{krr}, which first requires a definition of a \gls{rkhs} containing permissible \gls{rir} functions. Following the approach of \cite{brunnstromTimedomainsubmitted}, a \gls{rkhs} consisting of discrete-time sound fields will be constructed in this section. First, the \gls{dft} is described, to relate the time and frequency domain responses of a sound field. Then, a \gls{rkhs} for each frequency are aggregated using the \gls{dft} to obtain the desired function space. This section is a minimal restatement of definitions and results from \cite{brunnstromTimedomainsubmitted}, within which more details can be found.
	
	\subsection{Frequency-domain representation}\label{sec:frequency-domain-representation}
	The \gls{dft} is a transform between the set of time-domain sequences $\td = \real^{L}$, and frequency-domain sequences $\fd = \complex^{L_f-2} \times \real^2$ if $L$ is even and  $\fd = \complex^{L_f-1} \times \real$ if $L$ is odd, where $L_f = \lfloor\frac{ L}{2}\rfloor + 1$. The forward transform $\mathcal{F} : \td \rightarrow \fd$ and inverse transform $\mathcal{F}^{-1} : \fd \rightarrow \td$ between $\tilde{\bm{v}} \in \td$ and $\bm{v} \in \fd$ are defined as 
	\begin{equation}
		\begin{aligned}
			(\mathcal{F} \tilde{\bm{v}})_l &= \sum_{n=0}^{L-1} e^{2 \pi \imag n l  / L} (\tilde{\bm{v}})_n \quad 0 \leq l < L_f, \\
			(\mathcal{F}^{-1} \bm{v})_n &= \mathfrak{Re} \biggl[\sum_{l = 0}^{L_f-1} c_l e^{-2 \pi \imag n l / L} (\bm{v})_l \biggr] \quad 0 \leq n < L,  \\
		\end{aligned}
	\label{eq:dft-definitions}
	\end{equation}
	The real-part operator is defined as $\mathfrak{Re} [v] = \frac{1}{2} (v + \conj{v})$, and
	\begin{equation}
		c_l = \begin{cases}
			\frac{2}{L} & \quad \text{if } 0 < l < \frac{L}{2} \\
			\frac{1}{L} & \quad \text{if } l=0 \text{ or } l=\frac{L}{2}.
		\end{cases}
	\end{equation}
	It is convenient that the transform is unitary, like the standard \gls{dft} is with appropriate scaling. This is achieved by choosing the inner product of $\fd$ as 
	\begin{equation}
		\langle \bm{u}, \bm{v} \rangle_{\fd} = \mathfrak{Re}[\bm{v}^\herm \bm{C} \bm{u}]
		\label{eq:fd-inner-product}
	\end{equation}

	This transform is sometimes referred to as the \textit{real} \gls{dft}, as it assumes the time-domain signal to be real, and removes the redundancy in the form of conjugate symmetry in the frequency domain signal. The time-convention used in \eqref{eq:dft-definitions} is consistent with acoustics literature  \cite{martinMultiple2006}, but is opposite some popular implementations of the \gls{fft} \cite{bradburyJAX2018, harrisArray2020}.

	\subsection{Plane wave model}
	The sound field model is based on the Herglotz wave function for each of the sampled frequencies $\omega_l = \qty[parse-numbers=false]{\frac{2 \pi f_s l}{L}}{\radian / \second}$ in $\fd$, where $f_s$ is the sampling rate in \unit{\hertz}.  Any time-domain sound field $\tilde{\bm{u}} : \Omega \rightarrow \td$ can be represented as
	\begin{equation}
		\tilde{\bm{u}}(\bm{r}) = \mathcal{F}^{-1} \int_{\unitsphere} \bm{E}(\bm{r}, \unitvec{\bm{d}}) \pwcoeff{\bm{u}}(\unitvec{\bm{d}}) \,ds(\unitvec{\bm{d}}),
		\label{eq:frequency-domain-plane-wave-decomposition}
	\end{equation}
	where the directionality function $\pwcoeff{\bm{u}} : \unitsphere \rightarrow \fd$ represents the complex plane wave coefficients at the sampled frequencies. The set of permissible directionality functions is $L^2(\unitsphere \rightarrow \fd)$, which denotes the square integrable functions from the unit sphere $\unitsphere$ to $\fd$, where functions are considered equal if they only differ on a set with measure zero. The integration is performed with regards to the natural spherical measure $s$. The diagonal $\bm{E}(\bm{r}, \unitvec{\bm{d}}) \in \fdmatrix$ contains plane wave basis functions for a plane wave incoming from the direction $\unitvec{\bm{d}}$, and can be represented by the diagonal matrix
	\begin{equation}
		(\bm{E}(\bm{r}, \unitvec{\bm{d}}))_{ll} = \begin{cases}
			e^{-\imag \frac{\omega_l}{c}\bm{r}^\top \unitvec{\bm{d}} } & 0 \leq l < \frac{L}{2} \\
			\cos(\frac{\omega_l}{c}\bm{r}^\top \unitvec{\bm{d}}) & \text{if } l = \frac{L}{2}.
		\end{cases}
		\label{eq:plane-wave-sampled}
	\end{equation}
	where linear bounded operators with signature $\fd \rightarrow \fd$ are denoted by $\fdmatrix$. 



	\subsection{Reproducing kernel Hilbert space}
	The \gls{rkhs} of sound field functions can be defined in terms of the operator $\mathcal{A} : L^2(\unitsphere \rightarrow \fd) \rightarrow \rkhstime$, which takes a directionality function $\pwcoeff{\bm{u}}$ and transforms it into a sound field $\tilde{\bm{u}}$ according to \eqref{eq:frequency-domain-plane-wave-decomposition}. The \gls{rkhs} can then be defined as 
	\begin{equation}
		\tilde{\mathcal{H}} = \{\mathcal{A} \pwcoeff{\bm{u}} \vert \pwcoeff{\bm{u}} \in L^2(\unitsphere \rightarrow \fd)\},
		\label{eq:multi-freq-rkhs}
	\end{equation}
	with the inner product
	\begin{equation}
		\langle \tilde{\bm{u}}, \tilde{\bm{v}} \rangle_{\rkhstime} = \int_{\unitsphere} \langle \pwcoeff{\bm{u}}(\unitvec{\bm{d}}), \pwcoeff{\bm{v}}(\unitvec{\bm{d}}) \rangle_{\fd} \,ds(\unitvec{\bm{d}}).
	\end{equation}
	
	A \gls{rkhs} is characterized by its kernel function. The kernel function $\tilde{\Gamma} : \Omega \times \Omega \rightarrow \tdmatrix$  of $\rkhstime$ is
	\begin{equation}
		\tilde{\Gamma}(\bm{r}, \bm{r}') = \mathcal{F}^{-1} \Gamma(\bm{r}, \bm{r}') \mathcal{F}
		\label{eq:time-domain-kernel-function}
	\end{equation}
	where $\Gamma : \Omega \times \Omega \rightarrow \fdmatrix$ is represented by a diagonal matrix with elements
	\begin{equation}
		\begin{aligned}
			(\Gamma(\bm{r}, \bm{r}'))_{ll} &= 
			\begin{cases}
				\kappa_{\omega_l}(\bm{r}, \bm{r}') & 0 \leq l < \frac{L}{2} \\
				\frac{1}{2} (\kappa_{\omega_l}(\bm{r}, \bm{r}') + \kappa_{\omega_l}(\bm{r}, -\bm{r}') ) & l = \frac{L}{2},
			\end{cases}
		\end{aligned}
		\label{eq:multi-freq-rkhs-kernel}
	\end{equation}
	Finally, the single-frequency kernel $\kappa_{\omega} : \Omega \times \Omega \rightarrow \complex$ is 
	\begin{equation}
		\kappa_{\omega}(\bm{r}, \bm{r}') = j_0\Bigl(\frac{\omega}{c} \lVert \bm{r} - \bm{r}' \rVert_2 \Bigr)
		\label{eq:single-freq-kernel}
	\end{equation}
	where $j_0$ is the zeroth order spherical Bessel function of the first kind. The speed of sound is denoted by $c$, which is assumed to be constant in $\Omega$. 
	
	\section{Sound field estimation}\label{sec:sound-field-estimation}
	\subsection{Kernel ridge regression}\label{sec:krr}
	To estimate $\tilde{\bm{u}}$, an optimization problem in the form of \gls{krr} will be formulated.  The relationship between the function $\tilde{\bm{u}}$ and the measured data can be expressed as a linear measurement operator	$\mathcal{M}_n : \rkhstime \rightarrow \real$, defined as 
	\begin{equation}
		\mathcal{M}_n \tilde{\bm{u}} = \langle \tilde{\bm{u}}(\bm{r}_n), \bm{\phi}(n)\rangle_{\td},
		\label{eq:measurement-operator}
	\end{equation}
	which means the data can be expressed as $p(n) = \mathcal{M}_n \tilde{\bm{u}} + s(n)$. 

	The goal is to minimize the difference between the measured data $p(n)$ and what the measurement model \eqref{eq:data-model} predicts, while applying appropriate regularization. Such a \gls{krr} problem can be stated as 
	\begin{equation}
		\min_{\tilde{\bm{u}} \in \rkhstime} \sum_{n=0}^{N-1} (p(n) - \mathcal{M}_n \tilde{\bm{u}} )^2 + \lambda \lVert \tilde{\bm{u}} \rVert_{\rkhstime}^2
		\label{eq:krr}
	\end{equation}
	where $\lambda \in \real_{\geq 0}$ is a parameter controlling the strength of the regularization.

	According to the representer theorem \cite{diwaleGeneralized2018}, the optimal solution to a \gls{krr} problem can be stated in terms of the adjoints of the measurement operators, meaning that the optimal solution to \eqref{eq:krr} is
	\begin{equation}
		\tilde{\bm{u}}^{\text{opt}} = \sum_{n=0}^{N-1} \mathcal{M}_n^{*} a_n
		\label{eq:optimal-form}
	\end{equation}
	for a priori unknown parameters $a_n \in \real$, and the adjoints of the measurement operators $\mathcal{M}_n$ as defined in \eqref{eq:measurement-operator}. 
	
	When the measurement operator is a point-evaluation operator, as is the case with stationary microphones in \cite{brunnstromTimedomainsubmitted}, the adjoint coincides with the kernel function of the \gls{rkhs}. In this case with moving microphones, the measurement operators $\mathcal{M}_n$ are not point-evaluation operators, and so their adjoints are distinct from the kernel function. The adjoint $\mathcal{M}_n^{*} : \real \rightarrow \rkhstime$ is
	\begin{equation}
		\mathcal{M}_n^{*} a = \tilde{\Gamma}(\cdot, \bm{r}_n) \bm{\phi}(n) a
		\label{eq:adjoint-measurement}
	\end{equation}
	which mainly differs from the kernel function in that the source signal $\bm{\phi}$ is taken into account.
	
	
	\subsection{Solving kernel ridge regression}\label{sec:solving-krr}
	Inserting the form of the optimal estimate \eqref{eq:optimal-form} into the cost function \eqref{eq:krr} leads to a finite-dimensional convex optimization problem in terms of the parameters $a_n$. With the parameters and data combined into vectors as $\bm{a} = (a_0, \dots, a_{N-1}) \in \real^N$ and $\bm{p} = (p(0), \dots, p(N-1)) \in \real^N$, the optimization problem is
	\begin{equation}
		\min_{\bm{a} \in \real^N} \lVert \bm{p} - \bm{K} \bm{a} \rVert_{\real^N}^2 + \lambda \langle \bm{K} \bm{a}, \bm{a} \rangle_{\real^N}
		\label{eq:krr-finite-dimensional}
	\end{equation}
	The kernel matrix $\bm{K} \in \real^{N \times N}$ is a positive semi-definite matrix defined as 
	\begin{equation}
		\begin{aligned}
			(\bm{K})_{nm} &= \mathcal{M}_n \mathcal{M}_m^{*} \\
			&= \langle  \tilde{\Gamma}(\bm{r}_n, \bm{r}_m) \bm{\phi}(m), \bm{\phi}(n) \rangle_{\td}. 
		\end{aligned}
	\label{eq:kernel-matrix}
	\end{equation}
	The optimization problem in \eqref{eq:krr-finite-dimensional} is a regularized least squares problem, and hence has a unique optimum, which is 
	\begin{equation}
		\bm{a} = (\bm{K} + \lambda \bm{I})^{-1} \bm{p}
		\label{eq:optimal-parameter}
	\end{equation}
	To reconstruct the sound field $\tilde{\bm{u}}$ at an arbitrary point $\bm{r} \in \Omega$ from the parameters $\bm{a}$, the expression
	\begin{equation}
		\tilde{\bm{u}}(\bm{r}) = \sum_{n=0}^{N-1} \tilde{\Gamma}(\bm{r}, \bm{r}_n) \bm{\phi}(n) (\bm{a})_n
		\label{eq:optimal-estimate}
	\end{equation}
	is used. The estimate using this method is equivalent to the estimate obtained from the Bayesian approach taken in \cite{brunnstromBayesian2025} for omnidirectional microphones.

	\subsection{Kernel ridge regression including prior knowledge}
	In many cases there are more prior knowledge available than only the physical properties of sound. In a \gls{krr} context, such prior knowledge can be included by constructing a regularization function penalizing unlikely sound fields, as has been demonstrated in \cite{horiuchiKernel2021, brunnstromVariable2022, uenoDirectionally2021, ribeiroSound2024, brunnstromTimedomainsubmitted} particularly for prior knowledge on the direction of the sound field. The \gls{krr} problem for moving microphones considered in this paper admits a similar method to construct regularization penalties as in \cite{brunnstromTimedomainsubmitted}. 
	
	Instead of solving the \gls{krr} problem stated in \eqref{eq:krr}, the problem to be solved is
	\begin{equation}
		\min_{\tilde{\bm{u}} \in \rkhstime} \sum_{n=0}^{N-1} (p(n) - \mathcal{M}_n \tilde{\bm{u}})^2 + \lambda \lVert \linreg \tilde{\bm{u}} \rVert_{\mathcal{Z}}^2
		\label{eq:krr-regularized}
	\end{equation}
	where $\linreg : \rkhstime \rightarrow \mathcal{Z}$ is an invertible bounded linear operator, and $\mathcal{Z}$ is a Hilbert space. The operator $\linreg$ should be constructed such that sound fields that are likely to occur according to the prior knowledge are given a small norm. 
	
	Solving \eqref{eq:krr-regularized} follows the derivations and results from Section~\ref{sec:krr} and Section~\ref{sec:solving-krr} exactly, but replacing the kernel function $\tilde{\Gamma}$ with the modified kernel function $\tilde{\Gamma}_r : \Omega \times \Omega \rightarrow \tdmatrix$, which is defined as
	\begin{equation}
		\tilde{\Gamma}_r(\bm{r}, \bm{r}') = \Bigl((\linreg^{*} \linreg)^{-1} \tilde{\Gamma}(\cdot, \bm{r}')\Bigr)(\bm{r}). 
	\end{equation}

 	If there is prior knowledge available on the directionality of the sound field, a directional weighting can be used. Following the definitions of \cite[Section V.A]{brunnstromTimedomainsubmitted}, the regularization operator $\linreg$ is 
 	\begin{equation}
 		(\linreg \tilde{\bm{u}})(\bm{r}) = \mathcal{F}^{-1} \int_{\unitsphere} \bm{E}(\bm{r}, \unitvec{\bm{d}}) \bm{W}(\hat{\bm{d}}) \pwcoeff{\bm{u}}(\hat{\bm{d}}) ds(\hat{\bm{d}}),
 	\end{equation}
  	which is characterized by the weighting $\bm{W} : \unitsphere \rightarrow \fdmatrix$. The modified kernel function is then
	\begin{equation}
		\begin{aligned}
		\tilde{\Gamma}_r(\bm{r}, \bm{r}') &= \mathcal{F}^{-1} \Gamma_r(\bm{r}, \bm{r}') \mathcal{F} \\
		\Gamma_r(\bm{r}, \bm{r}') &= \int_{\unitsphere} \bm{E}(\bm{r}, \unitvec{\bm{d}}) (\bm{W}(\unitvec{\bm{d}})^{*} \bm{W}(\unitvec{\bm{d}}))^{-1} \bm{E}(-\bm{r}', \unitvec{\bm{d}}) \,ds(\unitvec{\bm{d}})
		\end{aligned}
	\end{equation}
	where it is used that $\bm{E}(\bm{r}', \unitvec{\bm{d}})^{*} = \bm{E}(-\bm{r}', \unitvec{\bm{d}})$. 
	
	A practical special case is when $\bm{W}$ is diagonal for all $\unitvec{\bm{d}} \in \unitsphere$, referred to as a frequency-wise directional weighting. In that case, the weighting can be defined by a scalar-valued positive function $\gamma_l : \unitsphere \rightarrow \real_{> 0}$ per frequency, as
	\begin{equation}
		\begin{aligned}
		\Bigl((\bm{W}(\unitvec{\bm{d}})^{*} \bm{W}(\unitvec{\bm{d}}))^{-1} \Bigr)_{ll} = \gamma_l(\unitvec{\bm{d}})
		\end{aligned}
		\label{eq:frequency-wise-weighting}
	\end{equation}
	With such a weighting, the frequency-domain kernel $\Gamma_r$ is still diagonal, with diagonal elements $(\Gamma_r(\bm{r}, \bm{r}'))_{ll} = \kappa_{\omega}^{\gamma}(\bm{r}, \bm{r}')$ which for all frequencies except $\omega_l$ where $l=\frac{L}{2}$ is
	\begin{equation}
		\kappa_{\omega}^{\gamma}(\bm{r}, \bm{r}') = \int_{\unitsphere} e^{-\imag \frac{\omega}{c}(\bm{r}- \bm{r}')^{\top} \unitvec{\bm{d}}} \gamma(\unitvec{\bm{d}}) \,ds(\unitvec{\bm{d}})
	\end{equation}
	Such a weighting is the type of weighting employed in the frequency domain \gls{krr} methods of \cite{horiuchiKernel2021, brunnstromVariable2022, ribeiroSound2024, ribeiroPhysicsconstrained2024}, hence the methods therein for finding appropriate weighting functions $\gamma_l$ can be directly applied. 
	
	Particularly simple is the kernel when the weighting is $\gamma_l(\unitvec{\bm{d}}) = e^{-\beta_l \unitvec{\bm{\eta}}_l^\top \unitvec{\bm{d}}}$, where $\unitvec{\bm{\eta}}_l \in \unitsphere$ is the direction of the weighting, and $\beta_l \in \real_{> 0}$ is the strength of the weighting \cite[Section V.A]{brunnstromTimedomainsubmitted}. Using this weighting in the \gls{krr} problem leads to a preference for estimates where the sound field propagates in the $\unitvec{\bm{\eta}}_l$ direction. The kernel can be computed as 
	\begin{equation}
		\begin{aligned}
		(\Gamma_r(\bm{r}, \bm{r}'))_{ll} &= j_0(\sqrt{\bm{\xi}_l^\top \bm{\xi}_l}) \\
		\bm{\xi}_l&= \frac{\omega_l}{c} (\bm{r} - \bm{r}') + \imag \beta_l \unitvec{\bm{\eta}}_l
		\end{aligned}
	\label{eq:vonmisesfisher-kernel}
	\end{equation}
	for $0 \geq l < \frac{L}{2}$, as it is assumed that $\beta_{\frac{L}{2}} = 0$ to ensure a closed-form solution. 
	



	\subsection{Computation of the kernel matrix}
	The most straightforward implementation of the expression in \eqref{eq:kernel-matrix} to compute $\bm{K}$ would require $N^2$ matrix-vector products, each with a computational cost proportional to $N^2$. That would quickly lead to an infeasibly large computational cost. However, the structure of $\bm{K}$ in \eqref{eq:kernel-matrix} reveals a cheaper way of computing the same matrix if a frequency-wise weighting is used, as described in \eqref{eq:frequency-wise-weighting}. 
	
	
	Expanding the definition of $\tilde{\Gamma}$ gives
	\begin{equation}
		\begin{aligned}
			(\bm{K})_{nm} &= \langle \Gamma(\bm{r}_n, \bm{r}_m) \bm{\phi}_f(m), \bm{\phi}_f(n) \rangle_{\fd}
		\end{aligned}
		\label{eq:kernel-matrix2}
	\end{equation}
	where $\bm{\phi}_f(n) = \mathcal{F} \bm{\phi}(n)$. Using the definition of the inner product in \eqref{eq:fd-inner-product}, and exploiting the fact that $\Gamma$ is diagonal, leads to
	\begin{equation}
		\begin{aligned}
			(\bm{K})_{nm} &= \sum_{l=0}^{L_f-1} c_l \mathfrak{Re}[\kappa_{\omega_l}^{\gamma}(\bm{r}_n, \bm{r}_m) \conj{(\bm{\phi}_f(n))_l}  (\bm{\phi}_f(m))_l ]. 
		\end{aligned}
		\label{eq:kernel-matrix-sum}
	\end{equation}
	The expression in \eqref{eq:kernel-matrix-sum} requires $L_f N^2$ evaluations of the scalar-valued kernel function to construct $\bm{K}$. Although the computational cost can still be high if the measurement time is long or the sampling rate is high, the cost is significantly reduced compared to the straightforward implementation.
	
	After constructing $\bm{K}$, the dominant computational cost is solving the linear system in \eqref{eq:optimal-parameter} for potentially very large $N$. Therefore, for large $N$, it can be beneficial to use large scale matrix-free solvers \cite{paigeLSQR1982}, as was demonstrated in \cite{verburgDynamic2025}.
	

	\section{Fast sound field estimation with random Fourier features}
	\newcommand{\rngvec}{\bm{\xi}}
	It is well known that the computational cost of kernel methods scale poorly with regards to the number of data points, due to the construction and inversion of a matrix with $N^2$ elements, which has an asymptotic computational cost of $N^3$ for most algorithms. For sound field estimation in the frequency-domain with stationary microphones \cite{uenoKernel2018} this number is equal to the number of microphones, leading to a small matrix and low computational cost. Using a moving microphone, the number is equal to the number of time-domain samples, leading to very large computational cost for higher sampling rates or longer measurement times. In this section, a structured approach to lower the computational cost is proposed, by using \gls{rff} as introduced in \cite{rahimiRandom2007}. Initially restricted to translation-invariant kernels, the method have been extended to more classes of kernels, with additional methods for reducing computational cost further \cite{liuRandom2022}. In this section, specifically the approach in \cite{rahimiRandom2007} is considered. 

	\subsection{Random Fourier features for stationary microphones}\label{sec:rff-stationary}
	The \gls{rff} approach will first be described for stationary microphones, using the single-frequency \gls{krr} problem \cite{uenoKernel2018}. Although the need for computational cost reduction is rarely needed with this method, it is a step towards the more useful moving microphone case. 
	
	The \gls{krr} optimization problem for stationary microphones is 
	\begin{equation}
		\min_{u_\omega \in \rkhs_\omega} \sum_{m=1}^{M} \lvert p_{\omega m} - u_\omega(\bm{r}_m) \rvert^2 + \lambda \lVert u_\omega \rVert_{\rkhs_\omega}^2
		\label{eq:single-freq-krr}
	\end{equation}
	where the data $p_{\omega m} \in \complex$ is the complex sound pressure at frequency $\omega$ recorded by microphone $m$ at position $\bm{r}_m \in \Omega$. The \gls{rkhs} $\rkhs_\omega$ contains functions $u_\omega : \Omega \rightarrow \complex$ that represent the complex sound pressure at frequency $\omega$, for which the kernel function is
	\begin{equation}
		\kappa_{\omega}^{\gamma}(\bm{r}, \bm{r}') = \int_{\unitsphere} e^{-\imag \frac{\omega}{c} (\bm{r} - \bm{r}')^\top \unitvec{\bm{d}}} \gamma(\unitvec{\bm{d}}) \,ds(\unitvec{\bm{d}})
	\end{equation}
	where $\gamma$ is a scalar weighting function, corresponding to a frequency-wise weighting as defined in \eqref{eq:frequency-wise-weighting}. 


	Central to \gls{rff} is Bochner's theorem, which states that any shift-invariant kernel function $\kappa : \real^3 \times \real^3 \rightarrow \complex$ is the Fourier transform of a probability measure $\mu : \real^3 \rightarrow \real$. In this case, 
	\begin{equation}
		\kappa_{\omega}^{\gamma}(\bm{r}, \bm{r}') = \int_{\real^{3}}  e^{-\imag (\bm{r} - \bm{r}')^\top \rngvec} \mu_\omega(\rngvec) \,d\rngvec,
		\label{eq:bochners-theorem}
	\end{equation}
	under the assumption that $\kappa_{\omega}^{\gamma}$ is scaled such that the total measure of $\mu_\omega$ is one.

	The implication of \eqref{eq:bochners-theorem} is that $e^{-\imag (\bm{r} - \bm{r}')^\top \rngvec}$ is an unbiased estimate of $\kappa_{\omega}^{\gamma}(\bm{r}, \bm{r}')$ if $\rngvec$ is sampled according to $\mu_\omega$. An estimate with less variance can be obtained by sampling more that one $\rngvec$, leading to the approximation
	\begin{equation}
		\kappa_{\omega}^{\gamma}(\bm{r}, \bm{r}') \approx \frac{1}{D} \sum_{d=1}^{D} e^{-\imag  (\bm{r} - \bm{r}')^\top \rngvec_d} = \langle \bm{z}_\omega(\bm{r}), \bm{z}_\omega(\bm{r}') \rangle_{\complex^D}
		\label{eq:kernel-approx}
	\end{equation}
	where $\rngvec_1, \dots, \rngvec_D$ is a list of the $D$ sampled vectors, and $\bm{z}_\omega(\bm{r}) = \frac{1}{\sqrt{D}} (e^{-\imag \bm{r}^\top \rngvec_1}, \dots, e^{-\imag \bm{r}^\top\rngvec_D}) \in \complex^D$ is the random basis vector.

%

	The probability measure $\mu_\omega$ that satisfies \eqref{eq:bochners-theorem} for $\kappa_{\omega}^{\gamma}$ is
	\begin{equation}
		\mu_\omega(\rngvec) = \delta\Bigl(\lVert \rngvec\rVert_{\real^3} - \frac{\omega}{c}\Bigr) \gamma(\rngvec) C_\gamma
		\label{eq:probability-measure}
	\end{equation}
	where $\delta$ is the continuous Dirac delta function, and $C_\gamma$ is a normalization constant ensuring that the total measure is 1, which will depend on the choice of $\gamma$. The derivation of \eqref{eq:probability-measure} is detailed in Appendix~\ref{sec:probability-measure}. The \gls{rff} basis can be identified as a basis of plane waves sampled proportionally to the weighting function $\gamma$. 
	
	

	Applying the representer theorem to \eqref{eq:single-freq-krr} results in
	\begin{equation}
		\min_{\bm{a} \in \complex^M} \lVert \bm{p}_\omega - \bm{K}_\omega \bm{a}_\omega \rVert_{\complex^M}^2 + \lambda \langle\bm{K}_\omega \bm{a}_\omega, \bm{a}_\omega \rangle_{\complex^M}
		\label{eq:freq-domain-dual-problem}
	\end{equation}
	where $(\bm{K}_\omega)_{m m'} = \kappa_{\omega}^{\gamma}(\bm{r}_m, \bm{r}_{m'})$, the parameter vector is $\bm{a}_\omega \in \complex^{M}$, and $\bm{p}_\omega = (p_{\omega 1}, \dots, p_{\omega M}) \in \complex^M$. Using the kernel approximation in \eqref{eq:kernel-approx} means that the kernel matrix can be approximated as $\bm{K}_\omega \approx \bm{Z}_\omega \bm{Z}_\omega^\herm$, where $\bm{Z}_\omega \in \complex^{M \times D}$ is 
	\begin{equation}
		\bm{Z}_\omega = \begin{bmatrix}
			\bm{z}_\omega(\bm{r}_0)^{\top}  \\ \vdots \\ \bm{z}_\omega(\bm{r}_M)^{\top} \\
		\end{bmatrix}
	\end{equation}


	The optimization problem \eqref{eq:freq-domain-dual-problem} can be restated in terms of the random basis parameter vector $\bm{b}_\omega \in \complex^D$, which is defined as $\bm{b}_\omega = \bm{Z}_\omega^\herm \bm{a}_\omega$, by inserting \eqref{eq:kernel-approx} into \eqref{eq:freq-domain-dual-problem}. The optimization problem is then
	\begin{equation}
		\min_{\bm{b}_\omega \in \complex^D} \lVert \bm{p}_\omega - \bm{Z}_\omega \bm{b}_\omega \rVert_{\complex^M}^2 + \lVert \bm{b}_\omega \rVert_{\complex^M}^2
	\end{equation}
	for which the optimal solution is 
	\begin{equation}
		\bm{b}_\omega = (\bm{Z}_\omega^\herm \bm{Z}_\omega + \lambda \bm{I})^{-1} \bm{Z}_\omega^\herm \bm{p}_\omega
	\end{equation}
	The sound field can be reconstructed using
	\begin{equation}
		u_\omega(\bm{r}) = \bm{z}_\omega(\bm{r})^\top \bm{b}_\omega
	\end{equation}

	\subsection{Random Fourier features for moving microphones}\label{sec:rff-moving}
	In this section, the \gls{rff} approach will be applied to the moving microphone case. The following derivations hold for a frequency-wise directional weighting as described in \eqref{eq:frequency-wise-weighting}. The random approximation for the scalar-valued kernel \eqref{eq:kernel-approx} can be inserted into the kernel expression \eqref{eq:time-domain-kernel-function} to obtain
	\begin{equation}
		\tilde{\Gamma}(\bm{r}, \bm{r}') \approx \mathcal{F}^{-1} \frac{1}{D} \sum_{d=1}^{D} \bm{E}(\bm{r}, \rngvec_d) \bm{E}(-\bm{r}', \rngvec_d) \mathcal{F}
		\label{eq:td-kernel-approx}
	\end{equation}
	which is the random approximation of the kernel for $\rkhstime$. Due to $\bm{E}$ being diagonal, it is possible to use different values of $D$ for different frequencies in the approximation \eqref{eq:td-kernel-approx}. However, for simplicity of exposition, the $D$ is assumed to be the same for all frequencies. 

	Using \eqref{eq:td-kernel-approx} to calculate $\bm{K}$ as defined in \eqref{eq:kernel-matrix} gives
	\begin{equation}
		\begin{aligned}
		(\bm{K})_{nm} &\approx \frac{1}{D}\sum_{d=1}^{D} \langle \mathcal{F}^{-1}  \bm{E}(\bm{r}_n, \rngvec_d) \bm{E}(-\bm{r}_m, \rngvec_d)  \mathcal{F} \bm{\phi}(m), \bm{\phi}(n) \rangle_{\td} \\
		&=\frac{1}{D} \sum_{d=1}^{D} \langle \bm{E}(-\bm{r}_m, \rngvec_d)  \bm{\phi}_f(m), \bm{E}(-\bm{r}_n, \rngvec_d) \bm{\phi}_f(n) \rangle_{\fd}.
	\end{aligned}
	\label{eq:kernel-matrix-approx}
	\end{equation}
	
	The random basis vector $\bm{v}_{d}(n) \in \fd$ can be defined as $\bm{v}_{d}(n) = \frac{1}{\sqrt{D}} \bm{E}(-\bm{r}_n, \rngvec_d) \bm{\phi}_f(n)$ which implies
	\begin{equation}
		(\bm{K})_{nm} \approx \sum_{d=1}^{D} \langle \bm{v}_{d}(n), \bm{v}_{d}(m) \rangle_{\fd}
		\label{eq:K-approx-new-basis}
	\end{equation}

	The expression \eqref{eq:K-approx-new-basis} can be simplified by making use of the isomorphism between $\fd$ and $\real^L$. The invertible transform $\mathcal{S} : \fd \rightarrow \real^L$ separates the real and imaginary parts, and is defined as 
	\begin{equation}
		\begin{aligned}
			\mathcal{S} \bm{b} = &(\sqrt{c_0} \mathfrak{Re}[b_0], \dots, \sqrt{c_{\frac{L}{2}}} \mathfrak{Re}[b_{\frac{L}{2}}], \\
			&\sqrt{c_1}, \mathfrak{Im}[b_1], \dots, \sqrt{c_{\frac{L}{2}-1}}, \mathfrak{Im}[b_{\frac{L}{2}-1}]), 
		\end{aligned}
	\end{equation}
	assuming even $L$, with the definition being analogous for odd $L$. The transform has the useful property $\langle\bm{a}, \bm{b} \rangle_{\fd} = \langle \mathcal{S} \bm{a}, \mathcal{S} \bm{b}  \rangle_{\real^L}$. Defining $\tilde{\bm{v}}_d(n) = \mathcal{S} \bm{v}_d(n)$, \eqref{eq:K-approx-new-basis} can be expressed as 
\begin{equation}
	(\bm{K})_{nm} \approx \sum_{d=1}^{D} \langle \tilde{\bm{v}}_{d}(n), \tilde{\bm{v}}_{d}(m)  \rangle_{\real^L} 
	\label{eq:K-approx-new-real-basis}
\end{equation}
	Collecting the real random basis vectors into a matrix $\bm{V} \in \real^{N \times LD}$ as
	\begin{equation}
		\bm{V} = \begin{bmatrix}
			\tilde{\bm{v}}_1(0)^\top & \dots & \tilde{\bm{v}}_D(0)^\top\\
			\vdots & \ddots & \vdots \\
			\tilde{\bm{v}}_1(N-1)^\top & \dots & \tilde{\bm{v}}_D(N-1)^\top\\
		\end{bmatrix}
	\end{equation}
	gives the concise approximation expression
	\begin{equation}
		\bm{K} \approx \bm{V} \bm{V}^{\top}
		\label{eq:K-approx-matrix}
	\end{equation}

	A new set of parameters $\bm{b}_d \in \real^L$ can be defined as the parameter vector $\bm{a}$ transformed by the new basis as $\bm{b}_d = \sum_{n=0}^{N-1} \tilde{\bm{v}}_d(n) a_n$. More concisely, it can be expressed as $\bm{b} = \bm{V}^\top \bm{a}$ for $\bm{b} = (\bm{b}_1, \dots, \bm{b}_D) \in \real^{LD}$. Then, inserting \eqref{eq:K-approx-matrix} into the finite-dimensional optimization problem \eqref{eq:krr-finite-dimensional} leads to
	\begin{equation}
		\begin{aligned}
			&\min_{\bm{b} \in \real^{LD}} \lVert \bm{p} - \bm{V} \bm{b} \rVert_{\real^N}^2 + \lambda \lVert \bm{b} \rVert_{\real^{LD}}^2
		\end{aligned}
	\label{eq:moving-mic-primal}
	\end{equation}
	for which the optimal solution is 
	\begin{equation}
		\bm{b} = (\bm{V}^\top \bm{V} + \lambda \bm{I})^{-1} \bm{V}^{\top} \bm{p}
	\end{equation}


	To reconstruct the sound field at an arbitrary point $\bm{r} \in \Omega$, the expression \eqref{eq:optimal-estimate} is used, but must be adapted to the new basis. Inserting the kernel approximation \eqref{eq:td-kernel-approx} into \eqref{eq:optimal-estimate}, the following is obtained
	\begin{equation}
		\begin{aligned}
			\tilde{\bm{u}}(\bm{r}) &\approx \frac{1}{D}\sum_{d=1}^D  \mathcal{F}^{-1} \bm{E}(\bm{r}, \rngvec_d) \sum_{n=0}^{N-1} \bm{v}_{d}(n) a_n \\
			&= \mathcal{F}^{-1} \frac{1}{\sqrt{D}} \sum_{d=1}^D  \bm{E}(\bm{r}, \rngvec_d) \mathcal{S}^{-1} \bm{b}_d \\
		\end{aligned}
		\label{eq:rff-moving-mic-reconstruction}
	\end{equation}
	
	The expression in \eqref{eq:rff-moving-mic-reconstruction} shows that each element in $\bm{b}$ has an interpretation as a plane wave coefficient. It becomes clearer if the reconstructed frequency domain sound field $\bm{u}(\bm{r}) = \mathcal{F} \tilde{\bm{u}}(\bm{r})$ is inspected, which is
	\begin{equation}
		\begin{aligned}
			(\bm{u}(\bm{r}))_l = \frac{1}{\sqrt{D}} \sum_{d=1}^{D} e^{-\imag \frac{\omega_{l}}{c} \bm{r}^{\top} \rngvec_d} (\mathcal{S}^{-1} \bm{b}_d)_{l}
		\end{aligned}
	\end{equation}
	for $0 \leq l < \frac{L}{2}$. It is clear that $(\mathcal{S}^{-1} \bm{b}_d)_{l}$ is the plane wave coefficient for the plane wave of direction $\rngvec_d$ and frequency $\omega_l$. 
	
	The \gls{rff} approach leads to a plane wave expansion method similar to \cite{verburgDynamic2025}, meaning that both \gls{rff} and \cite{verburgDynamic2025} can be viewed as approximations of \gls{krr} through the discretization of the Herglotz wave function \eqref{eq:frequency-domain-plane-wave-decomposition}. Their advantage is that the computational cost is dependent on the number of basis functions $LD$ rather than the number of samples $N$, which is preferable for long measurement times. However, in contrast to \cite{verburgDynamic2025}, the \gls{rff} approach naturally extends to new kernels, such as those obtained by a regularization operator $\linreg$ distinct from the identity operator, or those constructed for related sound field estimation problems \cite{ribeiroPhysicsconstrained2024}. In addition, approaches to further improve the effectiveness and reduce computational cost can be obtained from the many proposed extensions to the \gls{rff} scheme \cite{liuRandom2022}. 
	

	\section{Evaluation}\label{sec:evaluation}
	In this section the proposed \gls{krr} and \gls{rff} methods will be evaluated, and will be compared against existing sound field estimation methods with stationary microphones and moving microphones. Simulated data will be used to characterize the performance of the methods, followed by real data to validate the performance in practice.

 	The estimation performance will primarily be evaluated in terms of the \gls{nmse}, defined as 
	\begin{equation}
		\text{NMSE} = \frac{\sum_{\bm{r} \in \mathcal{E}} \lVert \tilde{\bm{u}}(\bm{r}) -\tilde{\bm{u}}^{\text{true}}(\bm{r}) \rVert_{\td}^2}{\sum_{\bm{r} \in \mathcal{E}}\Vert \tilde{\bm{u}}^{\text{true}}(\bm{r}) \rVert_{\td}^2},
	\end{equation}
	where $\mathcal{E} = \{\bm{r}_1, \dots, \bm{r}_E\}$ is the set of the $E$ evaluation points used in the experiment. In some cases the \gls{nmse} as a function of frequency is considered, in which the normalization is made frequency-wise, such that \gls{nmse} for frequency $\omega_l$ is
	\begin{equation}
		\text{NMSE}_{\omega_l} = \frac{\sum_{\bm{r} \in \mathcal{E}} \lvert (\bm{u}(\bm{r}))_l - (\bm{u}^{\text{true}}(\bm{r}))_l \rvert^2}{\sum_{\bm{r} \in \mathcal{E}}\lvert (\bm{u}^{\text{true}}(\bm{r}))_l \rvert^2}.
	\end{equation}
	where $\bm{u}(\bm{r}) = \mathcal{F} \tilde{\bm{u}}(\bm{r})$.

	\subsection{Methods}
	A number of sound field estimation methods will be compared, including the proposed \gls{krr} and \gls{rff}, using stationary or moving microphones. Following is a list of the methods in addition to the names used to refer to them. 
	

	\textbf{KRR-S}: Using stationary microphones, the \gls{krr} method using the unweighted kernel \eqref{eq:single-freq-kernel}. The method is often referred to as kernel interpolation with a diffuse kernel \cite{uenoKernel2018}. 
	
	\textbf{KRR-SD}: Using stationary microphones, the \gls{krr} method using a directional weighting as shown in \eqref{eq:vonmisesfisher-kernel}. The direction $\unitvec{\bm{\eta}}$ is chosen as the direction from the source to the center of the region of interest. 
	
	
	\textbf{KRR-M}: Using moving microphones, the method is the proposed method described in Section~\ref{sec:sound-field-estimation} without any weighting, i.e., $\linreg$ is chosen as the identity operator. This method is equivalent to the Bayesian method in \cite{brunnstromBayesian2025} with omnidirectional microphones. 
	
	\textbf{KRR-MD}: Using moving microphones, the method is the proposed method described in Section~\ref{sec:sound-field-estimation} using a frequency-wise directional weighting. The same weighting is used as KRR-SD. 
	
	
	\textbf{RFF-S}: Using stationary microphones, the \gls{rff} method proposed in Section~\ref{sec:rff-stationary}, using uniformly sampled basis directions. This method is the \gls{rff} approximation of KRR-S. 
	
	\textbf{RFF-SD}: Using stationary microphones, the \gls{rff} method proposed in Section~\ref{sec:rff-stationary}, using directionally informed sampling of the basis directions. This method is the \gls{rff} approximation of KRR-SD. The probability measure $\mu_\omega$ obtained from the chosen weighting function is the von Mises-Fisher distribution \cite[Section 9.3.2]{mardiaDirectional2000}, which can easily be sampled from \cite{jakobNumerically2015}. 
	
	
	\textbf{RFF-M}: Using moving microphones, the \gls{rff} method proposed in Section~\ref{sec:rff-moving} using uniformly sampled basis directions. This method is the \gls{rff} approximation of KRR-M. 
	
	\textbf{RFF-MD}: Using moving microphones, the \gls{rff} method proposed in Section~\ref{sec:rff-moving}, using directionally informed sampling of the basis directions. The same distribution is used as RFF-SD. This method is the \gls{rff} approximation of KRR-MD.

	\textbf{Spatial spectrum}: Using moving microphones, the method is based on a basis expansion using spherical wave functions \cite{katzbergSpherical2021}. The method can be viewed as a truncated version of the Bayesian method in \cite{brunnstromBayesian2025}. Alternatively, it can be viewed as a spherical wave function counterpart to the plane wave basis expansion methods of  \cite{verburgDynamic2025} and \gls{rff}. The method has a single additional parameter $r_\text{max}$, which is set according to the maximum radius of the moving microphone trajectory. 
	
	\textbf{Nearest neighbour}: Using stationary microphones, the estimated \gls{rir} for any position is chosen as the measured \gls{rir} from the closest microphone, without making any changes to it. This method serves as an indication of the difficulty of the chosen estimation problem, and is a baseline which all considered methods should improve upon.

	\subsection{Random Fourier features for stationary microphones}
	To begin with, the validity of the \gls{rff} method for stationary microphones will be demonstrated. To demonstrate the representation power of each method, the experiment is constructed to have very favourable conditions for sound field estimation. This also aligns with the moving microphone case where there are generally more data points compared to the number of basis functions. More realistic conditions will be considered for the moving microphone case, which is the focus of this paper. 
	\begin{figure}
		\centering
		\includegraphics{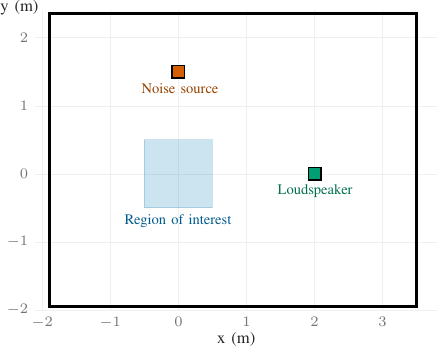}
		\label{fig:array-pos}
		\caption{The geometry of the simulated room, region of interest, loudspeaker and noise source. }
	\end{figure}
	
	\begin{figure}
		\includegraphics{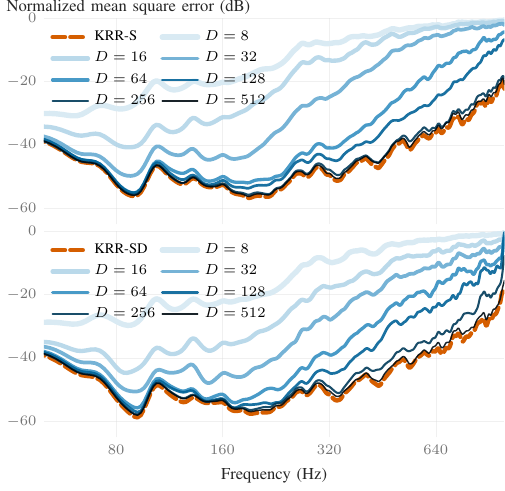}
		\caption{The estimation performance for different numbers of random features. The figure shows that a higher number of features are needed for higher frequencies. The non-weighted methods KRR-S and RFF-S are compared (top) and the directionally weighted methods KRR-SD and RFF-SD (bottom).}
		\label{fig:num_basis_functions}
	\end{figure}
	The experiment is performed using \glspl{rir} simulated using the image-source method \cite{allenImage1979, scheiblerPyroomacoustics2018} at a sampling rate of $f_s = \qty{2000}{\hertz}$. The \glspl{rir} are high-pass filtered at \qty{20}{\hertz} to simulate the low-frequency reduction in amplitude from a loudspeaker. The simulated cuboid room is of size \qtyproduct{5.4 x 4.3 x 3.2}{\meter}, within which the region of interest is a cuboid of size \qtyproduct{1 x 1 x 0.25}{\meter}. The geometry of the room and source is shown in Fig.~\ref{fig:array-pos}. 
	
	Favourable conditions for sound field estimation are made by randomly placing $128$ microphones with uniform distribution in the region of interest. The reverberation time is $\text{RT}_{60} = \qty{0.12}{\second}$. The \glspl{rir} at the microphone positions are used as data after transforming the \glspl{rir} into the frequency domain using a $1000$-point \gls{dft}. The regularization parameter is set to $\lambda = 10^{-4}$. Due to the randomness of the \gls{rff} method, the same experiments is performed 10 times with different random seeds, and the results shown are the mean scores.
	
	The estimation performance for different numbers of basis functions can be seen in Fig.~\ref{fig:num_basis_functions} as a function of frequency. The \gls{nmse} decrease as the number of basis functions increase, until it converges to the same error as the \gls{krr} estimate. \Gls{krr} consistently performs better than \gls{rff} for all frequencies, as expected. The experiment validates the proposal of using randomly sampled plane waves as an approximation of the kernel function in \eqref{eq:kernel-approx}.
	
	A higher number of basis functions are necessary for higher frequencies. Given that a higher number of basis functions increases computational cost, this implies that it could be effective to chose $D$ in a frequency-dependent manner. A comparison between the RFF-S and RFF-SD in Fig.~\ref{fig:num_basis_functions} indicates that the directional weighting increases estimation performance when using a lower number of basis functions. 

	\subsection{Random Fourier features for moving microphones}\label{fig:exp-rff-moving}
	\begin{figure}
		\includegraphics{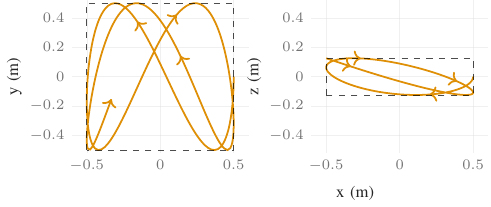}
		\caption{The Lissajous trajectory used for the moving microphone in simulations, shown projected onto a horizontal plane (left) and a vertical plane (right). }
		\label{fig:trajectory_rff_experiment}
	\end{figure}
	\begin{figure*}
		\includegraphics{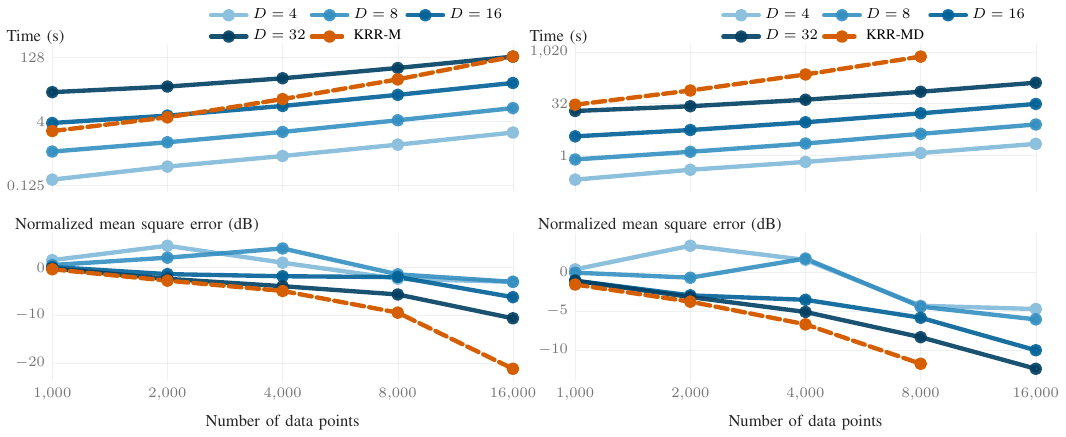}
		\caption{The clock time (top) and the \acrshort{nmse} (bottom) for RFF-M and KRR-M (left) and RFF-MD and KRR-MD (right). }
		\label{fig:sig_len_vs_cpu}
	\end{figure*}
	The \gls{rff} method is of more practical use for the moving microphone case, where the number of data points, and therefore the computational cost, of \gls{krr} becomes high. In this section the \gls{krr} and \gls{rff} methods will be compared on simulated data in terms of estimation performance and computational cost. Except where mentioned explicitly, the same simulation parameters and geometry is used as in Section~\ref{sec:rff-stationary}. A sampling rate of $f_s = \qty{1000}{\hertz}$ is used, and the \gls{rir} length is $L = 500$ samples, corresponding to \qty{0.5}{\second}. The microphone moves along a Lissajous curve, normalized to a speed of \qty{0.5}{\meter \per \second}, which is shown in Fig.~\ref{fig:trajectory_rff_experiment} for a measurement length of $N = 16000$. The regularization parameter is set to $\lambda = L \cdot 10^{-3}$.


	Estimates using RFF-M and RFF-MD are calculated for $D = 4, 8, 16$ and $32$ and compared against KRR-M and KRR-MD respectively. The results are shown in Fig.~\ref{fig:sig_len_vs_cpu}, where the estimates are calculated using only the first $N = 1000, 2000, 4000, 8000$, and $16000$ samples of data of the trajectory shown in Fig.~\ref{fig:trajectory_rff_experiment}. Due to the randomness of \gls{rff}, the results shown are the mean results from $10$ realizations. 

	To demonstrate the relationship in computational cost between \gls{krr} and \gls{rff}, the mean clock time of calculating the estimates are shown in Fig.~\ref{fig:sig_len_vs_cpu}. The code was executed on the CPU of a laptop with a AMD Ryzen 7 PRO 4750U processor. All methods are implemented in Python using JIT-compiled code in JAX \cite{bradburyJAX2018}, and are available at \cite{brunnstromAudio2025}. The computational cost of KRR-M and KRR-MD increases much faster compared to RFF-M and RFF-MD as the number of data points $N$ increase. Especially KRR-MD is costly, and is therefore not calculated for $N=16000$. In contrast, RFF-M and RFF-MD have essentially the same computational cost, that instead depends considerably on $D$. 
	
\begin{figure}
		\includegraphics{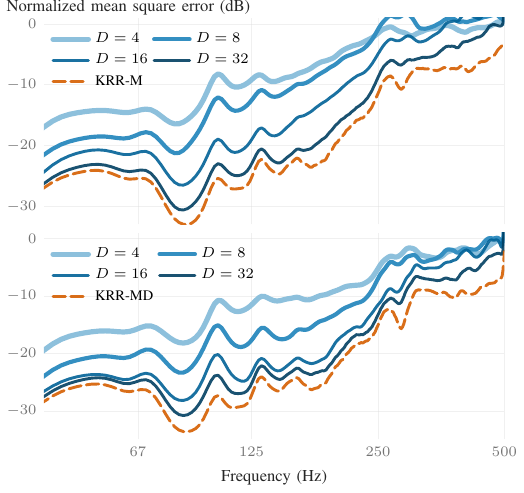}
		\caption{The estimation performance for different numbers of random features using moving microphones and $N = 8000$. The non-weighted methods KRR-M and RFF-M are compared (top) and the directionally weighted methods KRR-MD and RFF-MD (bottom). }
		\label{fig:num_basis_functions_moving}
	\end{figure}

	While \gls{krr} has higher computational cost for large $N$, it provides more accurate estimates for any given $N$. In Fig.~\ref{fig:sig_len_vs_cpu} the \gls{nmse} is shown, which demonstrates clearly that for a given amount of data, \gls{krr} obtains a lower \gls{nmse} than \gls{rff}. This means that there is a trade-off between estimation performance and computational cost that also depends on the amount of data available. Therefore, as a rule of thumb, \gls{rff} can be considered for long acquisition times. The \gls{nmse} in Fig.~\ref{fig:sig_len_vs_cpu} is calculated from the frequency bins between \qtyrange{20}{480}{\hertz}, because the source has little energy for very low frequencies, and the estimates very close to the Nyquist frequency are not reliable, as Fig.~\ref{fig:num_basis_functions_moving} shows.

	Finally, the estimation performance as a function of frequency is shown in Fig.~\ref{fig:num_basis_functions_moving} for $N=8000$. A similar pattern to RFF-S and RFF-D can be seen, where small values of $D$ are sufficient to estimate low frequencies well, but not higher frequencies. 
	

	
	
	\subsection{Estimation of simulated room impulse responses}\label{sec:experiment-sim-data}
	Using a directional weighting for moving microphones is a new feature of the proposed method, so it will be considered more closely in this section. As the effectiveness of the directional weighting chosen depends on the directionality of the sound field itself, the effect of the reverberation time will be investigated. In addition, the robustness of the different methods to the choice of regularization parameter will be shown. Except where explicitly mentioned, the same simulation parameters and geometry are used as in Section~\ref{fig:exp-rff-moving}. 
	
	The first $N=8000$ samples of the trajectory shown in Fig.~\ref{fig:trajectory_rff_experiment} is used as data for the moving microphone. The $M=16$ stationary microphones are placed in the center of each $L=500$ length segment of the trajectory. This choice leads to a fair comparison with the stationary microphone methods, as the spatial sampling is similar. A noise source is placed in the room at $(0, 1.5, 0) \unit{m}$, shown in Fig~\ref{fig:array-pos}, which emits white Gaussian noise at a power such that the microphone signal has an \gls{snr} of \qty{30}{\decibel}. The estimates are evaluated on an equally spaced grid of $E = 2000$ evaluation points within the region of interest, where the space between each point is \qty{0.05}{\meter}.

	\begin{figure}
		\includegraphics{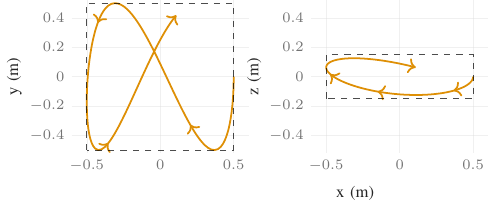}
		\caption{The 3-dimensional trajectory used in the experiment, shown projected onto a horizontal plane (left) and a vertical plane (right). }
		\label{fig:trajectory_dir_experiment}
	\end{figure}
	
\begin{figure}
		\includegraphics{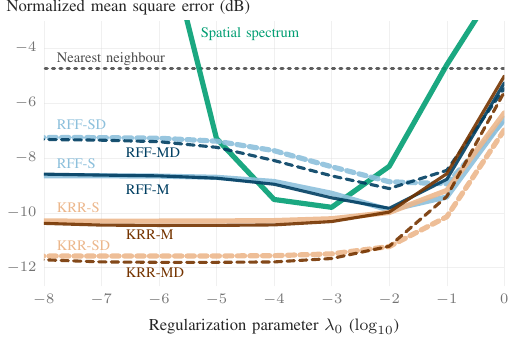}
		\caption{The estimation performance for all methods as a function of their regularization parameter, for the simulated data. }
		\label{fig:mse_per_reg_sim_data}
	\end{figure}

	
	All considered methods except nearest neighbour require a scalar regularization parameter to be set. Estimates are calculated using the regularization parameter $\lambda = L \lambda_0$ for KRR-M, KRR-MD, RFF-M, and RFF-MD, and using the regularization parameter $\lambda = \lambda_0$ for the other methods, where $\lambda_0 = 10^{t}$ for $t = (-8, -7, \dots, 0)$. A reverberation time of \qty{0.2}{\second} is used. The \gls{nmse} for all parameter values are shown in Fig.~\ref{fig:mse_per_reg_sim_data}. The results indicate a difference in robustness, where \gls{krr} degrades considerably less compared to \gls{rff} and spatial spectrum when a suboptimal $\lambda$ is chosen. All methods degrade significantly when $\lambda$ is chosen as $1$ or higher. 

	KRR-M is equivalent to the maximum a posteriori estimate of \cite{brunnstromBayesian2025} if the regularization parameter is chosen as $\lambda= \frac{L \sigma_s^2}{\sigma_{\alpha}^2}$, where $\sigma_s^2$ is the power of the noise $s$, and $\sigma_{\alpha}^2$ is the prior variance of the spherical wave function coefficients in \cite{brunnstromBayesian2025}. Even though the value of $\sigma_{\alpha}^2$ is not necessarily known, the expression suggests that a reasonable choice of the regularization parameter $\lambda$ is proportional to the inverse of the \gls{snr}, which is particularly helpful in environments with time-varying noise power levels. A similar connection exists between KRR-S and the Bayesian method in \cite{uenoSound2018}.

	\begin{figure}
\includegraphics{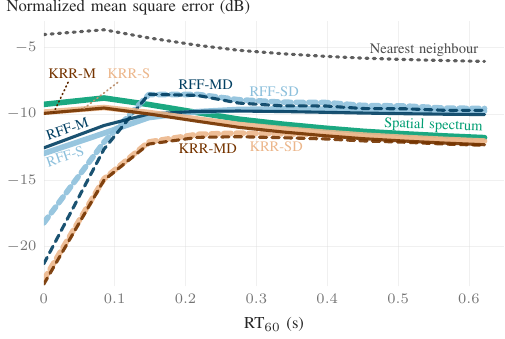}
		\caption{Estimation performance for varying simulated reverberation times. }
		\label{fig:mse_per_rt60_sim_data}
	\end{figure}
	To investigate the performance of the directional weighting, estimates are calculated for different reverberation times, for which the result is shown in Fig.~\ref{fig:mse_per_rt60_sim_data}. The reverberation times shown in Fig.\ref{fig:mse_per_rt60_sim_data} are the median $\text{RT}_{60}$ between the source and the evaluation points as measured using the Schroeder integration method \cite{schroederNew1965}. The directional parameter $\beta$ is chosen the same for all frequencies, by calculating estimates using KRR-SD for $32$ equally spaced values between 0 and 5, and using the $\beta$ which has the least \gls{nmse} on the evaluation points. 
	
	Fig.~\ref{fig:mse_per_rt60_sim_data} shows that the chosen directional weighting is particularly effective at short reverberation times. When the $\text{RT}_{60}$ is below \qty{0.1}{\second}, all directionally weighted methods have better performance compared to the non-weighted methods. As the reverberation time increases, the improvement for KRR-MD and KRR-SD becomes less pronounced, although their estimates remain more accurate than KRR-M and KRR-S. This indicates the value of the proposed approach, as the estimation performance is robustly improved, even with a very simple directional weighting. To improve performance further for long reverberation times, a more complicated weighting can be used, which more accurately takes into account the directionality of the early reflections and the late reverberation \cite{ribeiroSound2024}. 
	
	Fig.~\ref{fig:mse_per_rt60_sim_data} also shows that the accuracy of the estimates obtained from using stationary versus moving microphones are consistently very similar. While it is possible to find weighting functions using optimization methods such as \cite{ribeiroSound2024} for the proposed moving microphone method, finding such a function can be computationally costly, a cost which is made considerably larger by the large cost of the original moving microphone estimation problems. Given the close connection between \gls{krr} for moving and stationary microphones, this opens up the possibility of finding suitable weighting functions using stationary microphones. Such a procedure could be done using only a few measurement microphones, or a surrogate simulation that is close enough to the real room. The weighting functions can then be applied to the moving microphone data, giving a good trade-off between speed, computational cost, and estimation performance.


	\subsection{Estimation of real room impulse responses}\label{sec:real-rir-experiment}
	\begin{figure}
		\centering
		\includegraphics{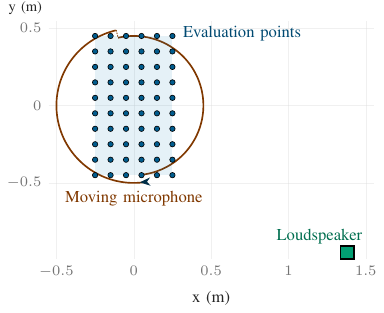}
		\caption{Trajectory for the moving microphone and positions of the evaluation points used in the experiment with real data from \cite{brunnstromExperimental2025}. }
		\label{fig:trajectory}
	\end{figure}
	Finally, the performance of the proposed method will be demonstrated on real \glspl{rir}. The dataset from \cite{brunnstromExperimental2025} is used, which has two microphones moving along circular trajectories of radii \qty{0.45}{\meter} and \qty{0.5}{\meter}. The dataset contains $E = 60$ evaluation points where \glspl{rir} are recorded using stationary microphones, which will be used as evaluation points. The trajectory, evaluation points, and source position are shown in Fig.~\ref{fig:trajectory}. Due to the similarity in performance between stationary and moving microphones demonstrated in section~\ref{sec:experiment-sim-data}, and the data available in the dataset, only moving microphone methods are considered in this section. The loudspeaker emits a periodic sweep with maximum frequency \qty{500}{\hertz}, which is sampled at a sampling rate of $f_s = \qty{1000}{\hertz}$. The \gls{snr} for the data used in the experiment is \qty{21}{\decibel}, which means that the data has more noise than the simulated data in Section~\ref{sec:experiment-sim-data}. 
	
\begin{figure}
		\includegraphics{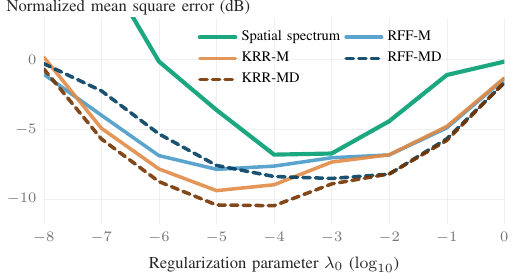}
		\caption{Estimation performance as a function of the regularization parameter for the real data. }
		\label{fig:mse_per_reg_real_data}
	\end{figure}

	The regularization parameter for the considered methods are once again investigated as described in Section~\ref{sec:experiment-sim-data}, but for the real data. The estimation accuracy as a function of the regularization parameter is shown in Fig.~\ref{fig:mse_per_reg_real_data}. The regularization parameter value giving the lowest \gls{nmse} in Fig.~\ref{fig:mse_per_reg_real_data} for each method is used for the remaining experiments. 
	
\begin{figure}
		\includegraphics{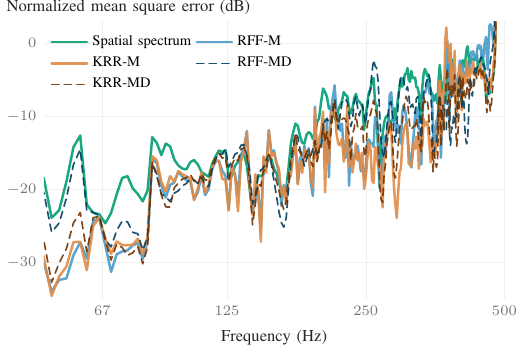}
		\caption{Estimation performance as a function of frequency using real data. }
		\label{fig:mse_per_freq_real_data}
	\end{figure}

	The mean \gls{nmse} for the considered methods is shown in Table~\ref{table:nmse-real-data}. The proposed KRR-MD has the lowest \gls{nmse}, indicating its applicability to real data. Both KRR-MD and RFF-MD has lower \gls{nmse} compared to their non-weighted counterparts KRR-M and RFF-M. Finally, KRR-M and KRR-MD both perform better than the three finite basis expansion methods RFF-M, RFF-MD and spatial spectrum.
	\begin{table}
		\centering
		\caption{Normalized mean square error in decibels for real data.}
		\label{table:nmse-real-data}
		\begin{tabular}{S[detect-weight, table-format=4.2, round-mode=places, round-precision=2] 
				S[detect-weight, table-format=4.2, round-mode=places, round-precision=2] S[detect-weight, table-format=4.2, round-mode=places, round-precision=2] 
				S[detect-weight, table-format=4.2, round-mode=places, round-precision=2]
				S[detect-weight, table-format=4.2, round-mode=places, round-precision=2]
			}
			\toprule 
			{KRR-M} & {KRR-MD} & {RFF-M} & {RFF-MD} & {Spatial spectrum} \\
			\midrule 
			-9.153066835706298  & \bfseries -10.232835126141342 &  -7.620921577418034  & -8.391146477856521 & -6.75971623274 \\
			\bottomrule
		\end{tabular}
	\end{table}
	The estimation performance as a function of frequency is shown in Fig.~\ref{fig:mse_per_freq_real_data}.


	
	\section{Conclusion}\label{sec:conclusion}
	A new method for sound field estimation using moving microphones have been proposed. The method is based on \gls{krr}, solving an optimization problem over a \gls{rkhs} with functions representing discrete time domain sound fields. The \gls{krr} method is shown to be more robust and produce more accurate estimates compared to methods that project the data onto a finite set of basis functions. One additional benefit over existing methods is that the proposed \gls{krr} method allows for prior information to be incorporated in the form of a regularization penalty. Using a directional weighting as regularization penalty has been shown to be effective, obtaining the lowest error on both simulated and real data. 
	
	\appendices
	
	\section{Adjoint of measurement operator}
	 The expression in \eqref{eq:adjoint-measurement} for the adjoint of the measurement operator $\mathcal{M}_n$ can be shown by the definition of the adjoint $\langle \mathcal{M}_n \tilde{\bm{u}}, a \rangle_{\real} = \langle \tilde{\bm{u}}, \mathcal{M}_n^{*} a \rangle_{\rkhstime}$. Expanding the left-hand side gives
	\begin{equation}
		\begin{aligned}
			\langle \mathcal{M}_n \tilde{\bm{u}}, a \rangle_{\real} &= 
			a \langle \tilde{\bm{u}}(\bm{r}_n), \bm{\phi}(n) \rangle_{\td} \\
			&= a  \int_{\unitsphere} \langle \mathcal{F}^{-1} \bm{E}(\bm{r}_n, \unitvec{\bm{d}}) \pwcoeff{\bm{u}}(\unitvec{\bm{d}}), \bm{\phi}(n) \rangle_{\td} \,ds(\unitvec{\bm{d}}) \\
			&=  \int_{\unitsphere} \langle \pwcoeff{\bm{u}}(\unitvec{\bm{d}}), a \bm{E}(-\bm{r}_n, \unitvec{\bm{d}}) \mathcal{F} \bm{\phi}(n) \rangle_{\fd} \,ds(\unitvec{\bm{d}}) \\
			&= \langle \tilde{\bm{u}}, \tilde{\bm{v}}_n \rangle_{\rkhstime},
		\end{aligned}
	\end{equation}
	where the directionality function of $\tilde{\bm{v}}_n = \mathcal{M}_n^{*} a$ is 
	\begin{equation}
		\pwcoeff{\bm{v}}_n = a \bm{E}(-\bm{r}_n, \unitvec{\bm{d}}) \mathcal{F} \bm{\phi}(n).
	\end{equation}
	Inserting the directionality function $\pwcoeff{\bm{v}}_n$ into the point evaluation expression in \eqref{eq:frequency-domain-plane-wave-decomposition} gives 
	\begin{equation}
		\begin{aligned}
		(\mathcal{M}_n^{*} a)(\bm{r}) &= a \mathcal{F}^{-1} \int_{\unitsphere} \bm{E}(\bm{r}, \unitvec{\bm{d}})\bm{E}(-\bm{r}_n, \unitvec{\bm{d}}) \mathcal{F} \bm{\phi}(n) \,ds(\unitvec{\bm{d}}) \\
		&= a \tilde{\Gamma}(\bm{r}, \bm{r}_n) \bm{\phi}(n),
		\end{aligned}
	\end{equation}
	which is the adjoint expression stated in \eqref{eq:adjoint-measurement}.
	
	\section{Fourier transform of frequency-domain kernel}\label{sec:probability-measure}
	
	The probability measure is derived for kernels with a directional weighting function $\gamma : \unitsphere \rightarrow \real$ \cite{koyamaSpatial2021}, which is defined as  
	\begin{equation}
		\kappa_{\omega}^{\gamma} (\bm{r}, \bm{r}')= \int_{\unitsphere} \gamma(\unitvec{\bm{d}}) e^{-\imag \frac{\omega}{c} (\bm{r} - \bm{r}')^{\top} \unitvec{\bm{d}}}\,ds(\unitvec{\bm{d}})
	\end{equation}
	Applying Bochner's theorem \eqref{eq:bochners-theorem}, the following is obtained
	\begin{equation}
		\int_{\unitsphere} \gamma(\unitvec{\bm{d}}) e^{-\imag \frac{\omega}{c} (\bm{r} - \bm{r}')^{\top} \unitvec{\bm{d}}} \,ds(\unitvec{\bm{d}}) = \int_{\real^3} \mu_\omega(\rngvec) e^{-\imag (\bm{r} - \bm{r}')^{\top} \rngvec} \,d\rngvec
	\end{equation}
	The two expressions are similar enough to determine $\mu_\omega$ by inspection. The equality holds if the measure $\mu_\omega$ is
		\begin{equation}
		\mu_\omega(\rngvec) = \delta(\lVert \rngvec\rVert_{\real^3} - \frac{\omega}{c}) \gamma(\unitvec{\bm{d}}) C_\gamma, 
	\end{equation}
	where $C_\gamma$ is a normalization constant such that $\mu_\omega$ is a probability measure. Because $\unitvec{\bm{d}}$ is a unit vector scaled by $\frac{\omega}{c}$, $\rngvec$ must always satisfy $\lVert \rngvec \rVert = \frac{\omega}{c}$. The implication is that with a directionally weighted kernel, the plane wave basis should be sampled with directions according to the weighting. 

	When the weighting is set to $\gamma(\unitvec{\bm{d}}) = 1$ for all $\unitvec{\bm{d}} \in \unitsphere$, then the kernel $\kappa_\omega$ is obtained, due to the identity
	\begin{equation}
		\int_{\unitsphere} e^{-\imag \bm{r}^{\top} \unitvec{\bm{d}}} \,ds(\unitvec{\bm{d}}) = j_0(\sqrt{\bm{r}^{\top} \bm{r}})
	\end{equation}
	for any complex $\bm{r} \in \complex^3$ \cite[Eq.65]{uenoDirectionally2021}. The associated probability measure is the uniform distribution on the sphere $\unitsphere$, defined explicitly as  
	\begin{equation}
		\mu_\omega(\rngvec) = \frac{1}{4 \pi \frac{\omega^2}{c^2}} \delta\Bigl(\lVert \rngvec\rVert_{\real^3} - \frac{\omega}{c}\Bigr). 
		\label{eq:uniform-on-sphere}
	\end{equation}

	\bibliographystyle{IEEEbib_mod}
	\bibliography{abbrev, refs}

\begin{thebibliography}{10}

\bibitem{betlehemPersonal2015}
T.~Betlehem, W.~Zhang, M.~A. Poletti, and T.~D. Abhayapala,
\newblock ``Personal sound zones: Delivering interface-free audio to multiple
  listeners,''
\newblock {\em IEEE Signal Process. Mag.}, vol. 32, no. 2, pp. 81--91, Mar.
  2015.

\bibitem{zhangSurround2017}
W.~Zhang, P.~N. Samarasinghe, H.~Chen, and T.~D. Abhayapala,
\newblock ``Surround by sound: A review of spatial audio recording and
  reproduction,''
\newblock {\em Appl. Sci.}, vol. 7, no. 5, May 2017,
\newblock {A}rt. no. 532.

\bibitem{ajdlerPlenacoustic2006}
T.~Ajdler, L.~Sbaiz, and M.~Vetterli,
\newblock ``The plenacoustic function and its sampling,''
\newblock {\em IEEE Trans. Signal Process.}, vol. 54, no. 10, pp. 3790--3804,
  Oct. 2006.

\bibitem{farinaSimultaneous2000}
A.~Farina,
\newblock ``Simultaneous measurement of impulse response and distortion with a
  swept-sine technique,''
\newblock in {\em Proc. {AES} Conv.} Feb. 2000.

\bibitem{mullerTransferfunction2001}
S.~M{\"u}ller and P.~Massarani,
\newblock ``Transfer-function measurement with sweeps,''
\newblock {\em J. Audio Eng. Soc.}, vol. 49, no. 6, pp. 443--471, June 2001.

\bibitem{antweilerNLMStype2008}
C.~Antweiler, A.~Telle, and P.~Vary,
\newblock ``{{NLMS-type}} system identification of {{MISO}} systems with
  shifted perfect sequences,''
\newblock in {\em Proc. Int. Workshop Acoust. Signal Enhancement ({IWAENC})},
  Sept. 2008.

\bibitem{nishidaRegionrestricted2022}
T.~Nishida, N.~Ueno, S.~Koyama, and H.~Saruwatari,
\newblock ``Region-restricted sensor placement based on gaussian process for
  sound field estimation,''
\newblock {\em IEEE Trans. Signal Process.}, vol. 70, pp. 1718--1733, 2022.

\bibitem{verburgOptimal2024}
S.~A. Verburg, F.~Elvander, T.~{van Waterschoot}, and E.~{Fernandez-Grande},
\newblock ``Optimal sensor placement for the spatial reconstruction of sound
  fields,''
\newblock {\em EURASIP J. Audio Speech Music Process.}, vol. 2024, no. 1, pp.
  41, Aug. 2024.

\bibitem{unnikrishnanSampling2013a}
J.~Unnikrishnan and M.~Vetterli,
\newblock ``Sampling high-dimensional bandlimited fields on low-dimensional
  manifolds,''
\newblock {\em IEEE Trans. Inf. Theory}, vol. 59, no. 4, pp. 2103--2127, Apr.
  2013.

\bibitem{grochenigMinimal2015}
K.~Gr{\"o}chenig, J.~L. Romero, J.~Unnikrishnan, and M.~Vetterli,
\newblock ``On minimal trajectories for mobile sampling of bandlimited
  fields,''
\newblock {\em Appl. Comput. Harmon. Anal.}, vol. 39, no. 3, pp. 487--510, Nov.
  2015.

\bibitem{katzbergCoherence2021}
F.~Katzberg, M.~Maass, and A.~Mertins,
\newblock ``Coherence based trajectory optimization for compressive sensing of
  sound fields,''
\newblock in {\em Proc. European Signal Process. Conf. ({EUSIPCO})}, Aug. 2021,
  pp. 116--120.

\bibitem{katzbergPositional2022}
F.~Katzberg, M.~Maass, R.~Pallenberg, and A.~Mertins,
\newblock ``Positional tracking of a moving microphone in reverberant scenes by
  applying perfect sequences to distributed loudspeakers,''
\newblock in {\em Proc. Int. Workshop Acoust. Signal Enhancement ({IWAENC})},
  Sept. 2022.

\bibitem{katzbergDoppler2024}
F.~Katzberg, M.~Maass, and A.~Mertins,
\newblock ``Doppler frequency analysis for sound-field sampling with moving
  microphones,''
\newblock {\em Front. Signal Process.}, vol. 4, Apr. 2024.

\bibitem{ajdlerDynamic2007}
T.~Ajdler, L.~Sbaiz, and M.~Vetterli,
\newblock ``Dynamic measurement of room impulse responses using a moving
  microphone,''
\newblock {\em J. Acoust. Soc. Am.}, vol. 122, no. 3, pp. 1636--1645, Sept.
  2007.

\bibitem{hahnComparison2016}
N.~Hahn and S.~Spors,
\newblock ``Comparison of continuous measurement techniques for spatial room
  impulse responses,''
\newblock in {\em Proc. European Signal Process. Conf. ({EUSIPCO})}, Aug. 2016,
  pp. 1638--1642.

\bibitem{hahnContinuous2017}
N.~Hahn and S.~Spors,
\newblock ``Continuous measurement of spatial room impulse responses using a
  non-uniformly moving microphone,''
\newblock in {\em Proc. {IEEE} Int. Workshop Appl. Signal Process. Audio
  Acoust. ({WASPAA})}, Oct. 2017, pp. 205--208.

\bibitem{hahnSimultaneous2018}
N.~Hahn and S.~Spors,
\newblock ``Simultaneous measurement of spatial room impulse responses from
  multiple sound sources using a continuously moving microphone,''
\newblock in {\em Proc. European Signal Process. Conf. ({EUSIPCO})}, Sept.
  2018, pp. 2180--2184.

\bibitem{antweilerPerfect1994}
C.~Antweiler and M.~D{\"o}rbecker,
\newblock ``Perfect sequence excitation of the {{NLMS}} algorithm and its
  application to acoustic echo control,''
\newblock {\em Ann. T{\'e}l{\'e}commun.}, vol. 49, no. 7, pp. 386--397, July
  1994.

\bibitem{katzbergSoundfield2017}
F.~Katzberg, R.~Mazur, M.~Maass, P.~Koch, and A.~Mertins,
\newblock ``Sound-field measurement with moving microphones,''
\newblock {\em J. Acoust. Soc. Am.}, vol. 141, no. 5, pp. 3220--3235, May 2017.

\bibitem{katzbergCompressed2018}
F.~Katzberg, R.~Mazur, M.~Maass, P.~Koch, and A.~Mertins,
\newblock ``A compressed sensing framework for dynamic sound-field
  measurements,''
\newblock {\em {IEEE/ACM} Trans. Audio, Speech, Lang. Process.}, vol. 26, no.
  11, pp. 1962--1975, Nov. 2018.

\bibitem{williamsFourier1999}
E.~G. Williams and J.~A. Mann,
\newblock {\em Fourier Acoustics: Sound Radiation and Nearfield Acoustical
  Holography}, vol. 108,
\newblock Academic Press, 1999.

\bibitem{katzbergSpherical2021}
F.~Katzberg, M.~Maass, and A.~Mertins,
\newblock ``Spherical harmonic representation for dynamic sound-field
  measurements,''
\newblock in {\em Proc. {IEEE} Int. Conf. Acoust., Speech, Signal Process.
  ({ICASSP})}, June 2021, pp. 426--430.

\bibitem{brunnstromBayesian2025}
J.~Brunnstr{\"o}m, M.~B. M{\o}ller, and M.~Moonen,
\newblock ``Bayesian sound field estimation using moving microphones,''
\newblock {\em IEEE Open J. Signal Process.}, vol. 6, pp. 312--322, Jan. 2025.

\bibitem{verburgDynamic2025}
S.~Verburg, Y.~Pene, and E.~{Fernandez-Grande},
\newblock ``Dynamic room impulse response measurements with a robotic arm,''
\newblock in {\em Proc. Forum Acusticum}, June 2025.

\bibitem{uenoSound2018}
N.~Ueno, S.~Koyama, and H.~Saruwatari,
\newblock ``Sound field recording using distributed microphones based on
  harmonic analysis of infinite order,''
\newblock {\em IEEE Signal Process. Lett.}, vol. 25, no. 1, pp. 135--139, Jan.
  2018.

\bibitem{brunnstromBayesian2024}
J.~Brunnstr{\"o}m, M.~B. M{\o}ller, J.~{\O}stergaard, and M.~Moonen,
\newblock ``Bayesian sound field estimation using uncertain data,''
\newblock in {\em Proc. Int. Workshop Acoust. Signal Enhancement ({IWAENC})},
  Sept. 2024, pp. 329--333.

\bibitem{brunnstromExperimental2025}
J.~Brunnstr{\"o}m, M.~B. M{\o}ller, T.~{van Waterschoot}, M.~Moonen, and
  J.~{\O}stergaard,
\newblock ``Experimental validation of sound field estimation methods using
  moving microphones,''
\newblock in {\em Proc. Forum Acusticum}, May 2025.

\bibitem{uenoKernel2018}
N.~Ueno, S.~Koyama, and H.~Saruwatari,
\newblock ``Kernel ridge regression with constraint of {{Helmholtz}} equation
  for sound field interpolation,''
\newblock in {\em Proc. Int. Workshop Acoust. Signal Enhancement ({IWAENC})}.
  Sept. 2018, pp. 436--440.

\bibitem{uenoDirectionally2021}
N.~Ueno, S.~Koyama, and H.~Saruwatari,
\newblock ``Directionally weighted wave field estimation exploiting prior
  information on source direction,''
\newblock {\em IEEE Trans. Signal Process.}, vol. 69, pp. 2383--2395, Apr.
  2021.

\bibitem{horiuchiKernel2021}
R.~Horiuchi, S.~Koyama, J.~G.~C. Ribeiro, N.~Ueno, and H.~Saruwatari,
\newblock ``Kernel learning for sound field estimation with l1 and l2
  regularizations,''
\newblock in {\em Proc. {IEEE} Int. Workshop Appl. Signal Process. Audio
  Acoust. ({WASPAA})}, Oct. 2021, pp. 261--265.

\bibitem{ribeiroSound2024}
J.~Ribeiro, S.~Koyama, and H.~Saruwatari,
\newblock ``Sound field estimation based on physics-constrained kernel
  interpolation adapted to environment,''
\newblock {\em {IEEE/ACM} Trans. Audio, Speech, Lang. Process.}, vol. 32, pp.
  4369--4383, Sept. 2024.

\bibitem{ribeiroPhysicsconstrained2024}
J.~G.~C. Ribeiro, S.~Koyama, and H.~Saruwatari,
\newblock ``Physics-constrained adaptive kernel interpolation for
  region-to-region acoustic transfer function: A {{Bayesian}} approach,''
\newblock {\em EURASIP J. Audio Speech Music Process.}, vol. 2024, Sept. 2024.

\bibitem{matsudaKernel2025}
R.~Matsuda, J.~Ribeiro, H.~Akiyama, and J.~Trevino,
\newblock ``Kernel ridge regression based sound field estimation using a rigid
  spherical microphone array,'' Aug. 2025,
\newblock arXiv preprint, arXiv:2508.03087.

\bibitem{brunnstromTimedomainsubmitted}
J.~Brunnstr{\"o}m, M.~Bo~M{\o}ller, J.~{\O}stergaard, S.~Koyama, T.~{van
  Waterschoot}, and M.~Moonen,
\newblock ``Time-domain sound field estimation using kernel ridge regression,''
  Sept. 2025,
\newblock arXiv preprint, arXiv:2509.05720.

\bibitem{rahimiRandom2007}
A.~Rahimi and B.~Recht,
\newblock ``Random features for large-scale kernel machines,''
\newblock {\em Adv. Neural Inf. Process. Syst.}, 2007.

\bibitem{liuRandom2022}
F.~Liu, X.~Huang, Y.~Chen, and J.~A.~K. Suykens,
\newblock ``Random features for kernel approximation: A survey on algorithms,
  theory, and beyond,''
\newblock {\em IEEE Trans. Pattern Anal. Mach. Intell.}, vol. 44, no. 10, pp.
  7128--7148, Oct. 2022.

\bibitem{brunnstromSpatial2025}
J.~Brunnstr{\"o}m, M.~B. M{\o}ller, J.~{\O}stergaard, T.~{van Waterschoot},
  M.~Moonen, and F.~Elvander,
\newblock ``Spatial covariance estimation for sound field reproduction using
  kernel ridge regression,''
\newblock in {\em Proc. European Signal Process. Conf. ({EUSIPCO})}, Sept.
  2025.

\bibitem{martinMultiple2006}
P.~A. Martin,
\newblock {\em Multiple Scattering: {{Interaction}} of Time-Harmonic Waves with
  {{N}} Obstacles}, vol. 107 of {\em Encyclopedia of Mathematics and Its
  Applications},
\newblock Cambridge University Press, 2006.

\bibitem{bradburyJAX2018}
J.~Bradbury, R.~Frostig, P.~Hawkins, M.~J. Johnson, C.~Leary, D.~Maclaurin,
  G.~Necula, A.~Paszke, J.~VanderPlas, S.~{Wanderman-Milne}, and Q.~Zhang,
\newblock ``{{JAX}}: {{Composable}} transformations of {{Python}}+{{NumPy}}
  programs,'' 2018,
\newblock [Online]. Available: github.com/jax-ml/jax.

\bibitem{harrisArray2020}
C.~R. Harris, K.~J. Millman, S.~J. {van der Walt}, R.~Gommers, P.~Virtanen,
  D.~Cournapeau, E.~Wieser, J.~Taylor, S.~Berg, N.~J. Smith, R.~Kern, M.~Picus,
  S.~Hoyer, M.~H. {van Kerkwijk}, M.~Brett, A.~Haldane, J.~F. {del R{\'i}o},
  M.~Wiebe, P.~Peterson, P.~{G{\'e}rard-Marchant}, K.~Sheppard, T.~Reddy,
  W.~Weckesser, H.~Abbasi, C.~Gohlke, and T.~E. Oliphant,
\newblock ``Array programming with {{NumPy}},''
\newblock {\em Nature}, vol. 585, no. 7825, pp. 357--362, Sept. 2020.

\bibitem{diwaleGeneralized2018}
S.~Diwale and C.~Jones,
\newblock ``A generalized representer theorem for {{Hilbert}} space-valued
  functions,'' Sept. 2018,
\newblock arXiv preprint, arXiv:1809.07347.

\bibitem{brunnstromVariable2022}
J.~Brunnstr{\"o}m, S.~Koyama, and M.~Moonen,
\newblock ``Variable span trade-off filter for sound zone control with kernel
  interpolation weighting,''
\newblock in {\em Proc. {IEEE} Int. Conf. Acoust., Speech, Signal Process.
  ({ICASSP})}, May 2022, pp. 1071--1075.

\bibitem{paigeLSQR1982}
C.~C. Paige and M.~A. Saunders,
\newblock ``{{LSQR}}: {{An}} algorithm for sparse linear equations and sparse
  least squares,''
\newblock {\em ACM Trans. Math. Softw.}, vol. 8, no. 1, pp. 43--71, Mar. 1982.

\bibitem{mardiaDirectional2000}
K.~Mardia and P.~Jupp,
\newblock {\em Directional Statistics},
\newblock Wiley Series in Probability and Statistics. John Wiley \& Sons, 2000.

\bibitem{jakobNumerically2015}
W.~Jakob,
\newblock ``Numerically stable sampling of the von {{Mises Fisher}}
  distribution on {{S2}} (and other tricks),''
\newblock Technical {{Report}}, Interactive Geometry Lab, ETH, June 2015,
\newblock [Online]. Available:
  mitsuba-renderer.org/{\textasciitilde}wenzel/files/vmf.pdf.

\bibitem{allenImage1979}
J.~B. Allen and D.~A. Berkley,
\newblock ``Image method for efficiently simulating small-room acoustics,''
\newblock {\em J. Acoust. Soc. Am.}, vol. 65, no. 4, pp. 943--950, Apr. 1979.

\bibitem{scheiblerPyroomacoustics2018}
R.~Scheibler, E.~Bezzam, and I.~Dokmani{\'c},
\newblock ``Pyroomacoustics: {{A Python}} package for audio room simulations
  and array processing algorithms,''
\newblock in {\em Proc. {IEEE} Int. Conf. Acoust., Speech, Signal Process.
  ({ICASSP})}, Apr. 2018, pp. 351--355.

\bibitem{brunnstromAudio2025}
J.~Brunnstr{\"o}m,
\newblock ``Audio signal processing collection (aspcol),'' 2025,
\newblock [Online]. Available: github.com/sounds-research/aspcol.

\bibitem{schroederNew1965}
M.~R. Schroeder,
\newblock ``New method of measuring reverberation time,''
\newblock {\em J. Acoust. Soc. Am.}, vol. 37, no. 3, pp. 409--412, Mar. 1965.

\bibitem{koyamaSpatial2021}
S.~Koyama, J.~Brunnstr{\"o}m, H.~Ito, N.~Ueno, and H.~Saruwatari,
\newblock ``Spatial active noise control based on kernel interpolation of sound
  field,''
\newblock {\em {IEEE/ACM} Trans. Audio, Speech, Lang. Process.}, vol. 29, pp.
  3052--3063, Aug. 2021.

\end{thebibliography}
	
\end{document}